\documentclass[10pt]{article}

\usepackage{amsmath}
\usepackage{amssymb}

\usepackage{graphicx}

\usepackage{cite}

\usepackage{color} 

\usepackage{enumerate}

\usepackage[labelfont=bf,labelsep=period,justification=raggedright]{caption}

\makeatletter
\renewcommand{\@biblabel}[1]{\quad#1.}
\makeatother

\date{}

\begin{document}

% Title must be 150 characters or less
\begin{flushleft}
{\Large
\textbf{Communication over the network of binary switches regulates the activation of A$_{2A}$ adenosine receptor}
}
% Insert Author names, affiliations and corresponding author email.
\\
Yoonji Lee$^{1}$, 
Sun Choi$^{1,\ast}$, 
Changbong Hyeon$^{2,\ast}$
\\
\bf{$^1$}National Leading Research Laboratory (NLRL) of Molecular Modeling and Drug Design, College of Pharmacy, Graduate School of Pharmaceutical Sciences, and Global Top 5 Research Program, Ewha Womans University, Seoul 120-750, Korea 
\\
\bf{$^2$}School of Computational Sciences, Korea Institute for Advanced Study, Seoul 130-722, Korea
\\
$^{\ast}$E-mail: sunchoi@ewha.ac.kr (S.C.); hyeoncb@kias.re.kr (C.H.)
\end{flushleft}

% Please keep the abstract between 250 and 300 words
\section*{Abstract}
Dynamics and functions of G-protein coupled receptors (GPCRs) are accurately regulated by the type of ligands that bind to the orthosteric or allosteric binding sites.
To glean the structural and dynamical origin of ligand-dependent modulation of GPCR activity, 
we performed total $\sim$ 5 $\mu$sec molecular dynamics simulations of A$_{2A}$ adenosine receptor (A$_{2A}$AR) in its apo, antagonist-bound, and agonist-bound forms in an explicit water and membrane environment, and examined the corresponding dynamics and correlation between the 10 key structural motifs that serve as the allosteric hotspots in intramolecular signaling network. 
We dubbed these 10 structural motifs ``binary switches" as they display molecular interactions that switch between two distinct states. 
By projecting the receptor dynamics on these binary switches that yield $2^{10}$ microstates, we show that 
(i) the receptors in apo, antagonist-bound, and agonist-bound states explore vastly different conformational space; 
(ii) among the three receptor states the apo state explores the broadest range of microstates; 
(iii) in the presence of the agonist, the active conformation is maintained through coherent couplings among the binary switches; and (iv) to be most specific, our analysis shows that W246, located deep inside the binding cleft, can serve as both an agonist sensor and actuator of ensuing intramolecular signaling for the receptor activation.
Finally, our analysis of multiple trajectories generated by inserting an agonist to the apo state underscores that the transition of the receptor from inactive to active form requires the disruption of ionic-lock in the DRY motif. 
%provides glimpses into early signatures of the inactive-to-active transition.  

% Please keep the Author Summary between 150 and 200 words
% Use first person. PLoS ONE authors please skip this step. 
% Author Summary not valid for PLoS ONE submissions.   
\section*{Author Summary}
As the key signal transmitters of a number of physiological processes, G-protein coupled receptors (GPCRs) are arguably one of the most important therapeutic targets. 
Orchestration of the intra-molecular signaling across transmembrane domain is key for the function of GPCRs. 
To investigate the microscopic underpinnings of intramolecular signaling that regulates the activation of GPCRs, we performed molecular dynamics simulations of the receptor in three distinct ligand-bound states using A$_{2A}$ adenosine receptor as a model system of GPCRs.
Statistical analyses on the dynamics of and correlation among the 10 ``binary switches" reveal that the three receptor states retain distinct dynamic properties. 
The antagonist- and agonist-bound forms of the receptors explore vastly different conformational space, and the apo form lies between them, yet located closer to the antagonist-bound form. 
In regard to the agonist-binding triggered activation mechanism, the correlation map among the 10 binary switches unequivocally shows that direct sensing of agonist ligand by the indole ring of W246 actuates the rest of intramolecular signaling.  

\section*{Introduction}
G-protein coupled receptors (GPCRs) are one of the most versatile membrane proteins that mediate cellular responses to a myriad of extracellular signals associated with our perception, cardiovasicular, and immune functions \cite{Rosenbaum2009Nature}. 
GPCRs relay extracellular signals to the cytoplasmic domain. 
For a given extracellular signal, there is a corresponding GPCR subtype that processes the incoming signal to the cellular downstream \cite{jaakola2008Science}. 
Consisting of seven transmembrane (TM) helices, each of which is connected to the next TM helix by either an intracellular loop (ICL) or an extracellular loop (ECL), the interior of GPCR forms an interhelical residue-to-residue interaction network that can transmit the signal specific to the ligand and/or receptor subtype. 

Binding of an agonist to the orthosteric site leads to conformational rearrangement of TM helices, transforming the inactive conformation to the active one, which in turn enables accommodation of G-proteins and intracellular signal transductions \cite{Rosenbaum2009Nature,rasmussen2011Nature,lebon2011Nature}. 
It is widely appreciated that together with highly conserved residues in the TM region, the activation of the receptors belonging to the class A GPCR family is regulated by a set of fingerprint residues called ``microswitches" \cite{Rosenbaum2009Nature,nygaard2009TPS}, which commonly include three structural motifs: DRY, CWxP, and NPxxY motifs (``x" denotes any amino acid residue). 
Upon binding of an agonist, the microswitch residues change the orientation of their side chain, which transforms the global configuration of TM helices into the active form and helps the intracellular domain accommodate G-proteins \cite{nygaard2009TPS}. 
In contast, binding of an inverse-agonist suppresses the GPCR function below the basal level, to which the apo or antagonist-bound forms of GPCRs is likely to be tuned \cite{Rosenbaum2009Nature}.

The allostery, a long-range communication between two remote sites \cite{Monod63JMB,edward2001PNAS,Gandhi08PNAS,Weinkam12PNAS}, is one of the key determinants for functions in many biomolecules. And it is of particlular interest for GPCRs because most of dynamic processes associated with GPCR activation and suppression involve signaling between extracellular and intracellular domains, established across the TM domain. 
Although the term ``allosteric modulation" is often strictly distinguished from the orthosteric signaling in GPCR community, 
both can be considered allosteric signaling in that their signalings are both physically long-ranged.   
While high-resolution crystal structures in the agonist-bound or antagonist-bound states provide unprecedented view of ligand-dependent modulation of GPCR conformation, minor conformational difference between distinct receptor states make it difficult to glean the microscopic origin of intramolecular signaling and modulation of GPCR activity.  
Unlike molecular machines \cite{Hyeon06PNAS,Horovitz05COSB,Hyeon11BJ,Vale00Science} and enzymes that exhibit large conformational changes \cite{Whitford07JMB,Lee11JACS}, to which one can apply a method such as normal mode analysis and its variation \cite{Bahar10ARB,Zheng06PNAS}, it is not straightforward to study the allosteric dynamics of biomolecules like GPCRs\cite{cui2008ProtSci,popovych2006NSMB,balabin2009PNAS,Lee14Proteins} whose structural changes between the inactive and active forms are relatively small (RMSD between inactive and active forms are only 1.78 \AA\ for A$_{2A}$ adenosine receptor; and 2.96 \AA\ for $\beta_2$ adrenergic receptor whose active state structure is crystalized with G$_s$-protein).
It is noteworthy that there is a growing realization that a certain class of proteins can display allosteric responses by modulating conformational fluctuations, but with little conformational change \cite{Cooper84EBJ,buchli2013PNAS,Motlagh2014Nature,Itoh10PNAS}. 
Global conformational changes themselves are not the sole physical origin of protein allostery. 
Thus, to gain a better understanding of allostery or long-range intramolecular signaling of GPCR, it is imperative to probe the dynamical features of key structural components and their correlations.    
Although experimental methods, such as fluorescence resonance energy transfer (FRET), and site-directed spin labeling (SDSL), nuclear magnetic resonance (NMR) \cite{Yao2006NCB, Hoffmann2008BJP, Ambrosio2011Neuropharmacol, Altenbach1996Biochemistry, Altenbach1999Biochemistry}, are useful to monitor the conformational changes of specific key residues, it is generally difficult to simultaneously probe the multiple sites of molecule and elucidate dynamic origin leading to allosteric molecular responses. 
Much effort has recently been made to study the mechanism of GPCR activation via molecular dynamics (MD) simulations \cite{Niesen11JACS,kohlhoff2014NatureChem,Dror2011PNAS, Li2013JACS,Pang2013Proteins,Lee2013CPL} by exploring the dynamic characteristics of GPCR conformations; however, systematic studies of cross-correlations among key structural motifs including microswtiches have not been carried out.

From sequence analysis, structural and biochemical studies using mutagenesis,
there is a general consensus for the class A GPCRs that 18 hotspot residues, called microswitches, are responsible for the activation mechanism \cite{Rosenbaum2009Nature,rasmussen2011Nature}.  
Recently we have shown that when GPCR structures are represented by a network of inter-residue contacts, many of those microswitches \cite{Rosenbaum2009Nature,nygaard2009TPS} retain high betweenness centrality values \cite{Lee14Proteins}. 
According to the network theory, vertices with high betweenness centralities in a given network are the sites that mediate flow of signal over the network \cite{freeman1979SocialNetworks,Goh01PRL}. 
A removal or alteration of such vertices from the network, which can be realized in the form of deletion \cite{delSol2006MSB} or more practically mutation into glycine \cite{Lee14Proteins}, could impair the network communication and be deleterious to intramolecular signaling. 
Among the 18 microswitches, our network analysis in Ref.\cite{Lee14Proteins} could identify 11 of them in A$_{2A}$ adenosine receptor (A$_{2A}$AR), and also showed that the majority of intra-molecular signaling pathways connecting the highly correlated residues between extra and intra-cellular domains pass through the 11 microswitches.
However, our previous study provided only a static view of intramolecular allosteric wiring. 
Thus, to better understand the intramolecular signaling of A$_{2A}$AR, 
it would be of great interest to directly probe the dynamics associated with these microswitches and their cross-correlations.
 
In this study, we performed each of 1 $\mu$sec all-atom molecular dynamics simulation of A$_{2A}$AR in apo, antagonist-bound (ZM-241385),\cite{dall2003BJP,jaakola2008Science} and agonist-bound (UK-432097) \cite{xu2011Science} forms (see Methods); and monitored the local dynamics of the microswitches.   
Based on the dynamics around the 11 microswitches, each of which displays molecular interactions that switches between two distinct states, we defined 10 ON/OFF binary switches as the microscopic components to describe the interhelical dynamics (note that the two microswitch residues of ionic-lock compose one on/off switch). 
Describing the dynamics of the receptor by employing 10 microscopic components amounts to projecting the entire dynamics into 10 local variables and considering $2^{10}$ microstates.  
By projecting the receptor dynamics on $2^{10}$ microstates, 
we show how the dynamics of each ligand-dependent macrostate differs from each other. 
Our simulations and analysis show that compared to that of the apo and antagonist-bound forms, the 10 binary switches of agonist-bound form retain greater cross-correlation and coherence, which is consistent with the notion that GPCR in agonist-bound form has a functional fidelity to transmitting its intra-molecular signals across the TM domain \cite{Lee14Proteins}. 
Detailed knowledge on the local dynamics and their dynamic correlation elucidated in this study for A$_{2A}$ adenosine receptors could be useful for the discovery of effective drugs.  

% Results and Discussion can be combined.
%\section*{Results}
%\section*{Discussion}

\section*{Results}
Among the 18 microswitches of A$_{2A}$AR, 
our previous study calculating the betweenness centrality of residue interaction network suggested that 11 of them are the putatively important spots for the intramolecular signal transmission \cite{Lee14Proteins}, which include N24$^{1.50}$, D52$^{2.50}$, D101$^{3.49}$, R102$^{3.50}$, W129$^{4.50}$, Y197$^{5.58}$, E228$^{6.30}$, W246$^{6.48}$, N284$^{7.49}$, P285$^{7.50}$, Y288$^{7.53}$ where the superscripts refer to the Ballesteros Weinstein numbering system \cite{ballesteros1995MethodsNeurosci}. 
Examining the local dynamics around the eleven allosteric hotspot residues \cite{Lee14Proteins} at different receptor states, we have defined 10 binary (on/off) switches. Below we first explain how we have defined each switch to be binary, which maximally separates the dynamic features of antagonist-bound state from those of agonist-bound state. \\
%Representation of GPCR conformations using the 10 binary switches, which amounts to projecting GPCR conformations onto $2^{10}$ microstates, enables us to demonstrate how distinct each  receptor state of GPCR is and how this difference modulate the GPCR allostery. \\

\noindent{\bf Ten binary switches defined from eleven hotspot residues for intra-molecular signaling.}

\begin{figure}[h]
\centering
 \includegraphics[width=0.9\columnwidth]{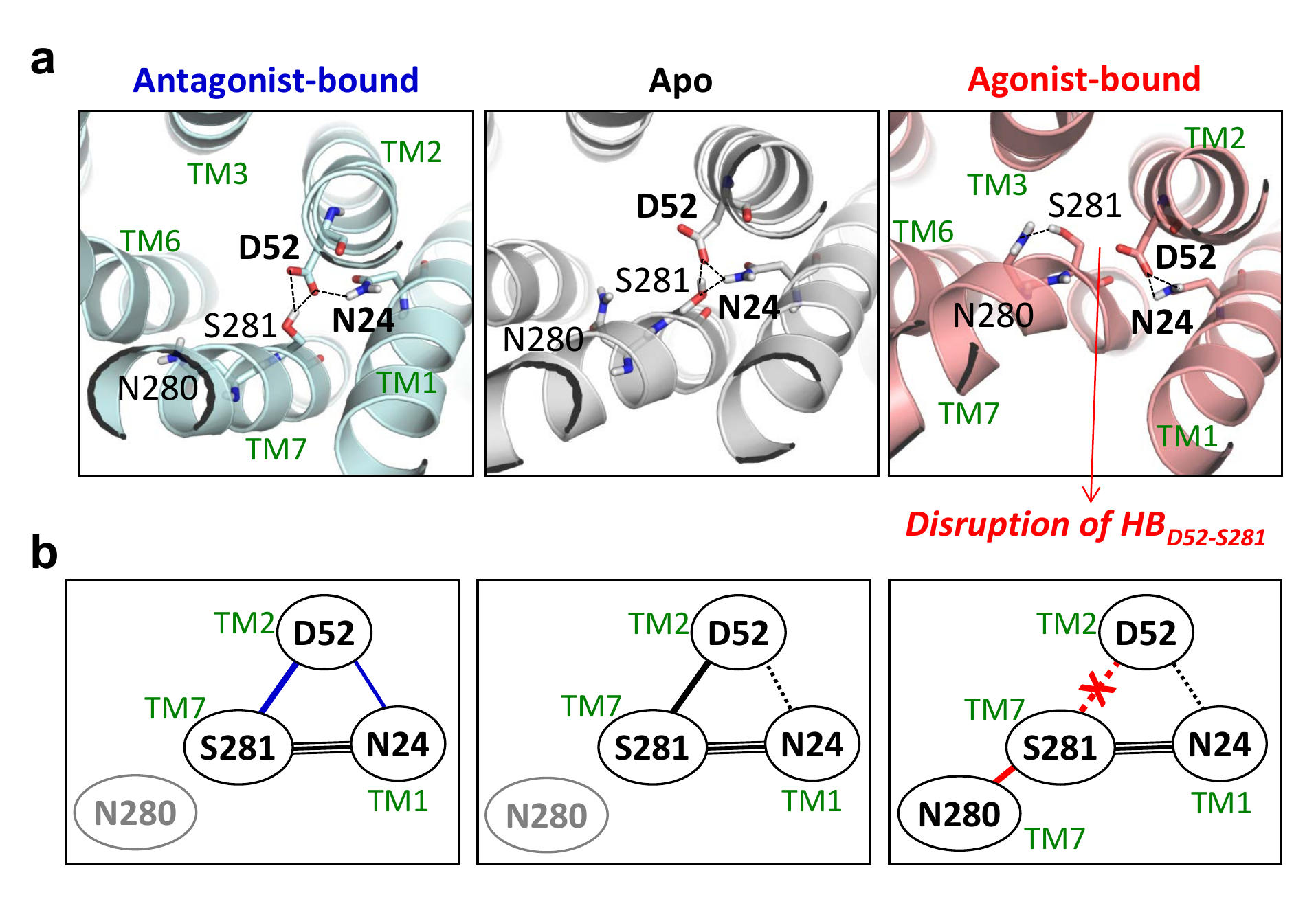}
  \caption{{\bf $\mathcal{S}1$ and $\mathcal{S}2$ at the interface between TM1, TM2, and TM7  for the antagonist-bound, apo, and agonist-bound forms.} (a) H-bonding network of the microswitch residues. (b) Diagram of the H-bond network in the three receptor states.
\label{TM1-2}}
\end{figure}

{\it Microswitches at the interfaces between TM1, TM2, and TM7 (Switches 1 \& 2)}:
In the class A GPCR family N24 and D52 are the most conserved microswitch residues in TM1 and TM2, respectively, with high betweenness centrality values \cite{Lee14Proteins}. 
Comparison of the H-bond network between three different  receptor states reveals notable differences in the receptor configuration around them (Fig.~\ref{TM1-2} and Fig.~S1).
First, in the apo form, H-bonds of N24 and D52 with S281$^{7.46}$ contribute to the stable conformation of TM helices. 
Binding of antagonist further stabilizes this conformation by incorporating an additional H-bond between N24 and D52. 
By contrast, binding of an agonist leads to the disruption of the H-bond between D52 and S281 (HB$_{D52-S281}$) and forms a new H-bond between S281 and N280$^{7.45}$, which is responsible for helix bending in TM7 (see below).  
Ligand-dependent switching dynamics of the H-bond of S281 from one side to the other is reminiscent of the salt-bridge switching, which is often observed at the interface between subunits in multisubunit proteins \cite{Hyeon06PNAS,Mono65JMB}. 
Hence, we designate HB$_{\text{N24-D52}}$ and HB$_{\text{D52-S281}}$ as the switch 1 ($\mathcal{S}1$) and the switch 2 ($\mathcal{S}2$), respectively.

\begin{figure}[h]
\centering
  \includegraphics[width=0.8\columnwidth]{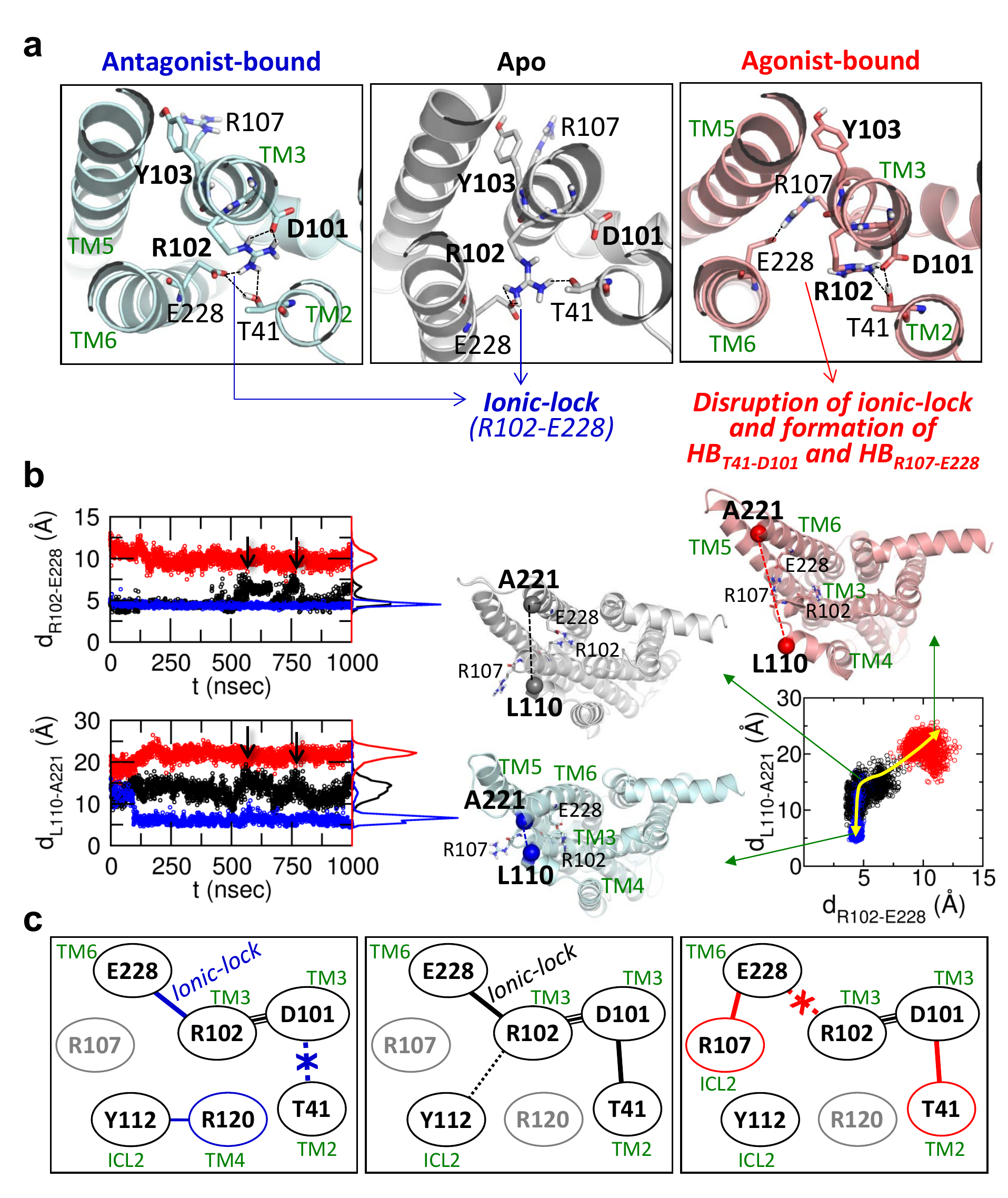}
  \caption{{\bf $\mathcal{S}3$ and $\mathcal{S}4$ defined in the DRY motif and ionic-lock.} (a) Configurations of the DRY motif and ionic-lock in the three receptor states. 
(b) The distances of R102$_{C_{\zeta}}$-E228$_{C_{\delta}}$ (left, top) and L110$_{C_{\alpha}}$-A221$_{C_{\alpha}}$ (left, bottom) are colored in black, blue, and red for the apo, antagonist-bound, and agonist-bound states, respectively, and their histograms are shown on the right side of the plots. 
On the right panel, scattered plot using the distances of R102-E228 and L110-A221 is shown with the yellow arrow depicting the conformational transition between three ligand forms. 
(c) Diagram of the H-bond network in the three receptor states. 
\label{DRY_motif}}
\end{figure}

{\it DRY motif and ionic-lock (Switches 3 \& 4)}:
The orientation of the TM5-ICL3-TM6 relative to the ICL2 is the hallmark of GPCR activation as the $10^o$ tilt of TM5-ICL3-TM6 enables the G-protein accommodation. 
Among others, the ``ionic-lock", the salt-bridge between R102 and E228, which keeps the TM3 and TM6 in close proximity, is considered the key structural motif that directly regulates the orientation of TM6 helix \cite{weis2008COSB}.  
Our simulations find that, in the antagonist-bound state, the ionic-lock is intact ($d_{\text{R102}_{C_{\zeta}}\text{-E228}_{C_{\delta}}}\approx 4$ \AA) (Fig.~\ref{DRY_motif}a and the top panel of Fig.~\ref{DRY_motif}b); while it is rarely formed in the agonist-bound state ($d_{\text{R102-E228}}\approx 10$ \AA). 
In the apo state, the ionic-lock maintains the identical distance with that of the antagonist-bound state, but occasionally breaks ($d_{\text{R102}_{C_
{\zeta}}\text{-E228}_{C_{\delta}}}\approx 7$ \AA) and reforms ($d_{\text{R102}_{C_{\zeta}}\text{-E228}_{C_{\delta}}}\approx 4$ \AA) (Fig.~\ref{DRY_motif} and Fig. S2).

The ionic-lock and the inter-residue distance ($d_{\text{L110-A221}}$) between L110$_{C_{\alpha}}$ in ICL2 and A221$_{C_{\alpha}}$ in ICL3 were simultaneously probed in Fig.~\ref{DRY_motif}b. 
It highlights the correlation between the ionic-lock and the orientation of TM5-ICL3-TM6 relative to TM3.
Dependence of $d_{\text{L110-A221}}$ on the  receptor state is clearly observed; $d_{\text{L110-A221}}\approx 22$ \AA\ (agonist-bound), 7 \AA\ (antagonist-bound), and 14 \AA\ (apo), which quantifies the position of TM6 helix relative to TM3 helix.   
For the apo form, the distance between ICL2 and ICL3 varies concomitantly with the state of ionic-lock (see the time traces after 500 nsec of Fig.~\ref{DRY_motif}b). 
The scatter plot of $d_{\text{R102-E228}}$ and $d_{\text{L110-A221}}$ (right panel of Fig.~\ref{DRY_motif}b) also shows a clear correlation between the dynamics of ionic-lock and the orientation of TM5-ICL3-TM6 domain. 
The status of ionic-lock affects the H-bond network around the DRY motif (Fig.~\ref{DRY_motif}c and Fig. S3). 
In the agonist-bound form, E228 forms the H-bond with R107 instead of R102.
Disruption of ionic-lock leads to release of TM6 from TM3, but as a counterbalance a new H-bond (HB$_{\text{T41-D101}}$) binds TM3 tightly with TM2.  
As HB$_{\text{T41-D101}}$ and the ionic-lock in the DRY motif can be used to discern the ligand-dependent macrostate of GPCR, we define HB$_{\text{T41-D101}}$ and the ionic-lock as $\mathcal{S}3$ and $\mathcal{S}4$, respectively.

\begin{figure}[h]
\centering
  \includegraphics[width=0.8\columnwidth]{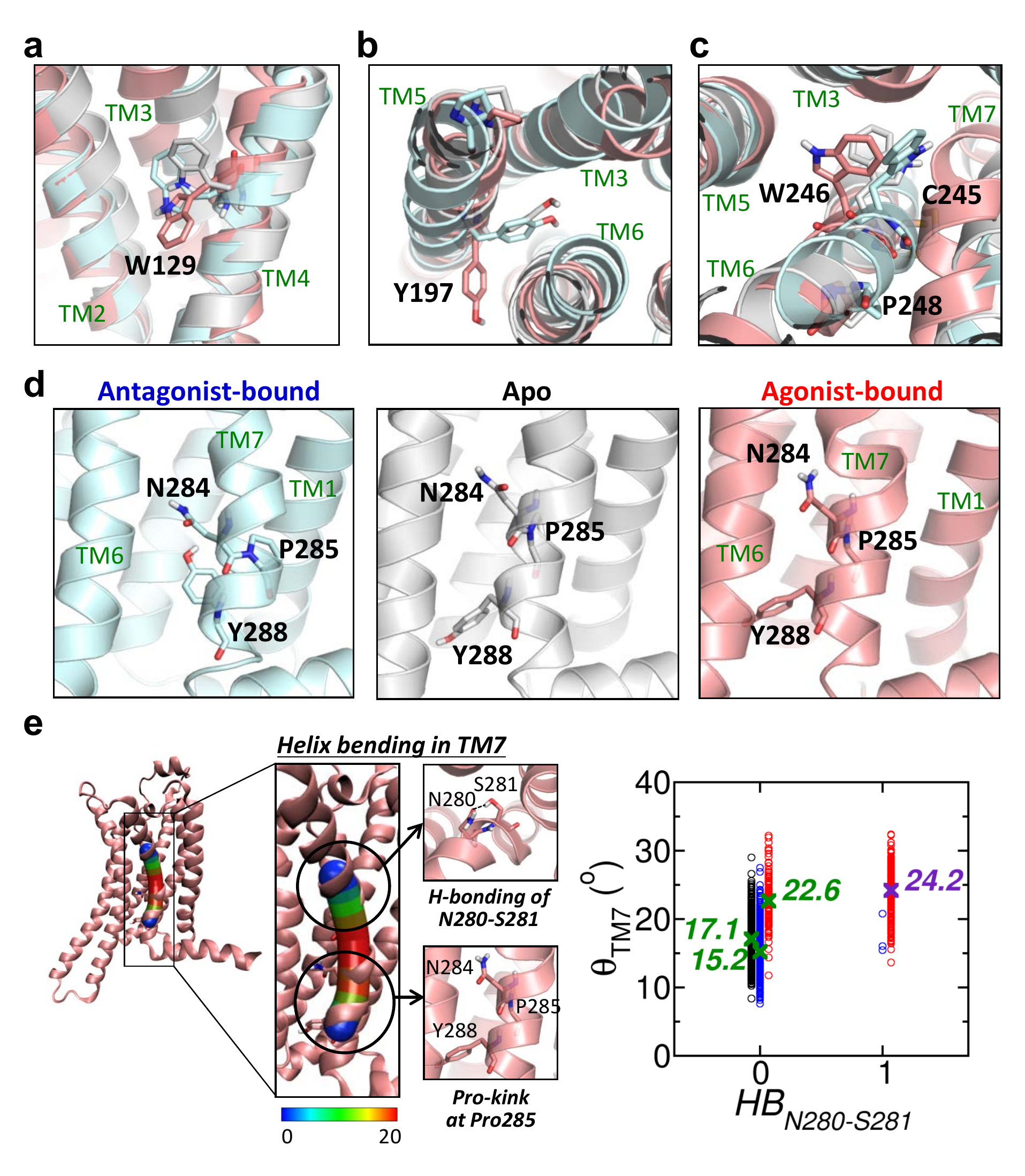}
  \caption{{\bf $\mathcal{S}5$ to $\mathcal{S}10$ defined from the rotameric switches in TM4, TM5, TM6, and TM7.} Rotameric states of (a) W129, (b) Y197, (c) CWxP motif, and (d) NPxxY motif are compared for the apo (white), antagonist-bound (cyan), and agonist-bound (pink) forms. (e) Helix bending in TM7. The helix bending angle (bendix) of TM7 was calculated using bendix program \cite{Dahl12Bioinformatics}. The helix is displayed as a cylinder marked with the heatmap ranging from 0$^o$ to 20$^o$. 
The scatter plot on right side depicts the relationship between H-bond of N280-S281 and the bending angle of the TM7 helix (apo : black, antagonist-bound: blue, agonist-bound states: red). 
The average bending angles are annotated with the symbols, X.
 \label{rotamers}}
\end{figure}

{\it Rotameric microswitches (Switches 5, 6, 7, 8, \& 10)}: 
Ligand-dependent microstates of the five residues W129, Y197, W246, N284, and Y288 are best characterized by the rotameric angles of their side chains (Fig. S4), which allow us to define these residues as $\mathcal{S}5$, $\mathcal{S}6$, $\mathcal{S}7$, $\mathcal{S}8$, and $\mathcal{S}10$, respectively:   
(i) In the agonist-bound state, W129 ($\mathcal{S}5$) shows asymmetric bimodal distribution of dihedral angle $P(\chi_2)$ that has a major peak at $\approx -20^o$ and a minor peak at $\approx 90^o$. 
In the apo and antagonist-bound forms, $P(\chi_2)$ is unimodal with a peak at $90^o$ (Fig.~\ref{rotamers}a and Fig. S5a); 
(ii) The dihedral angles of Y197 ($\mathcal{S}6$) also show distinct distribution in the three  receptor states (Fig.~\ref{rotamers}b and Fig. S5b);
(iii) W246 ($\mathcal{S}7$), known as a central switch located at the bottom of the orthosteric binding cleft \cite{nygaard2009TPS}, shows discrete transition between $\approx +90^o$ and $\approx -90^o$ with the receptor state-depedent angle value and frequency (Fig.~\ref{rotamers}c, Fig. S5c, and supporting movie M1);
(iv) N284 ($\mathcal{S}8$) and Y288 ($\mathcal{S}10$) in the NPxxY motif show receptor-state dependent dihedral angle distribution (Fig.~\ref{rotamers}d and Fig. S5d). 
Meanwhile, the dihedral angle distributions of P285, which also belongs to NPxxY motif, are little affected by the receptor state. 
Thus, we use the kink angle of P285 instead of its dihedral angle for defining the switch (see below).

{\it Helix bending induced by proline kink (Switch 9)}:
The geometry of proline residue constrains the backbone conformation and cause a sharp kink in alpha helix structures.  
Calculation of the helix bending angle \cite{Dahl12Bioinformatics} shows that in TM7 helix, the greatest kink is formed around P285, in particular, in the agonist-bound form (Fig.~\ref{rotamers}e). 
Comparison between structures for different receptor  states suggests that the H-bonding between N280$^{7.45}$ and S281$^{7.46}$, in the agonist-bound state, contributes to the helix bending in TM7. 
The scatter plot in Fig.~\ref{rotamers}e indicates that there is a direct correlation between TM7-helix bending ($\theta_{\text{TM7}}$) and HB$_{\text{N280-S281}}$. 
\\

\begin{figure*}[h]
\centering
  \includegraphics[width=1.0\columnwidth]{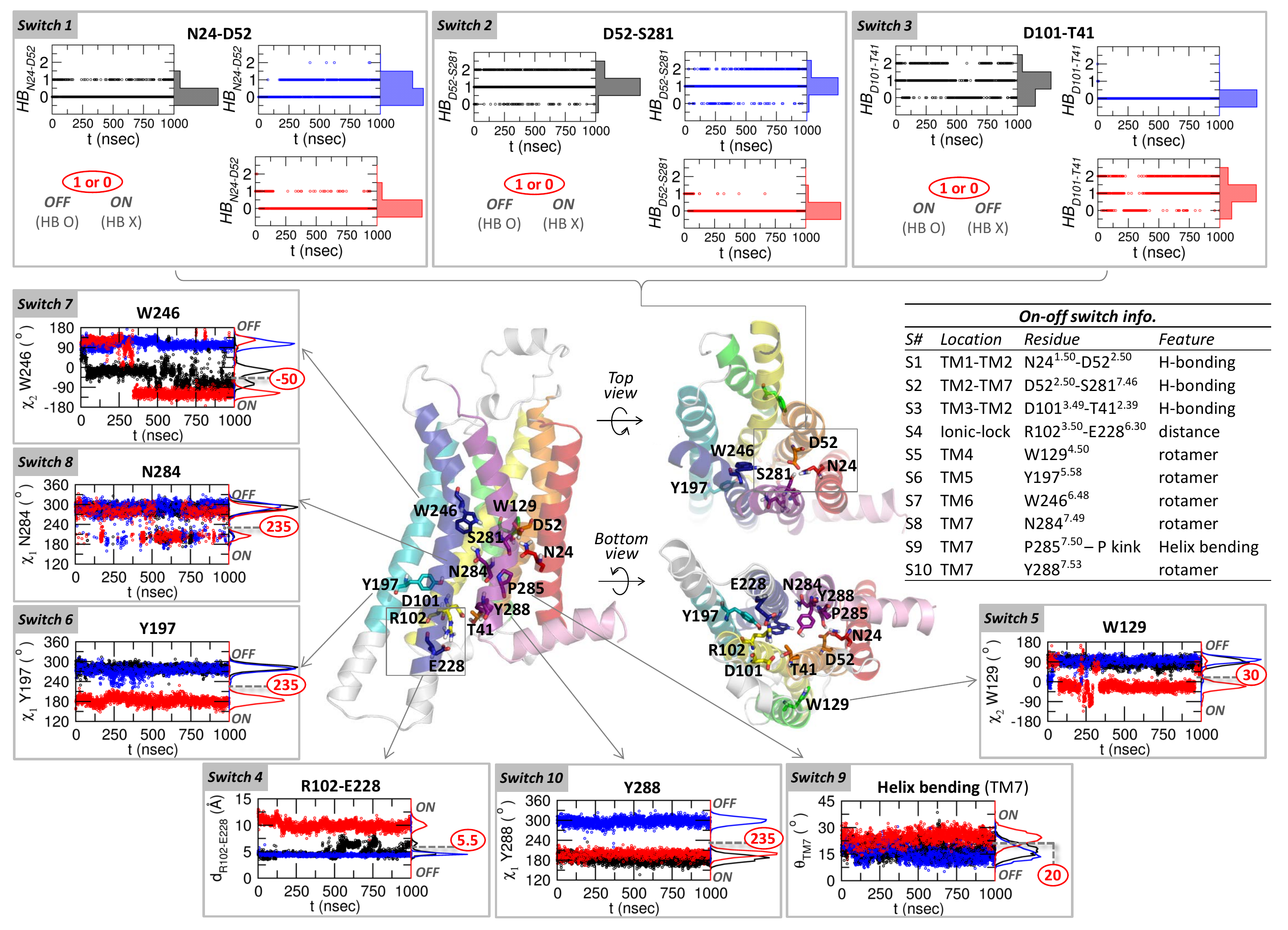}
  \caption{{\bf Ten binary switches.} The time traces of the apo, antagonist-bound, and agonist-bound forms are colored by black, blue, and red, respectively, and their histograms are shown on the right side of the panels. 
  From $\emph{S}4$ to $\emph{S}10$, the values separating the on and off states are marked in red circles. 
   \label{10_switches}}
\end{figure*}

\noindent{\bf Projection of GPCR dynamics onto 10-binary switches.}
The dynamic features of the 11 microswitch residues that display two-state switch-like molecular interactions allow us to define 10 binary switches (Fig.\ref{10_switches}). We choose the value of the $i$-th switch ($s_i$) as 1 or 0 (ON or OFF) in reference to the switch state in the agonist-bound form, so that the ``similarity" of each switch can be assessed in reference to the agonist-bound form. 
For example, the HB$_{\text{D52-S281}}$, corresponding to $\mathcal{S}$2, is disrupted in the agonist-bound form. 
In this case we set $s_2=1$ for disrupted H-bond and $s_2=0$ for intact H-bond. 
Also, for $\mathcal{S}7$ whose configuration is best described using rotameric angle, we consider $\mathcal{S}7$ to be in the ON state if the dihedral angle (C$_{\alpha}$-C$_{\beta}$-C$_{\gamma}$-C$_{\delta 1}$) of W246 (see Fig. S4) is equal or less than $-50^o$; otherwise, it is in the OFF state ($s_7=0$). 
Our 10 switch representation resembles the strategy in studying protein folding problem using ``correctness" of the configuration of each residue with respect to the native state \cite{Zwanzig95PNAS}.

\begin{figure}[h]
\centering
  \includegraphics[width=1.0\columnwidth]{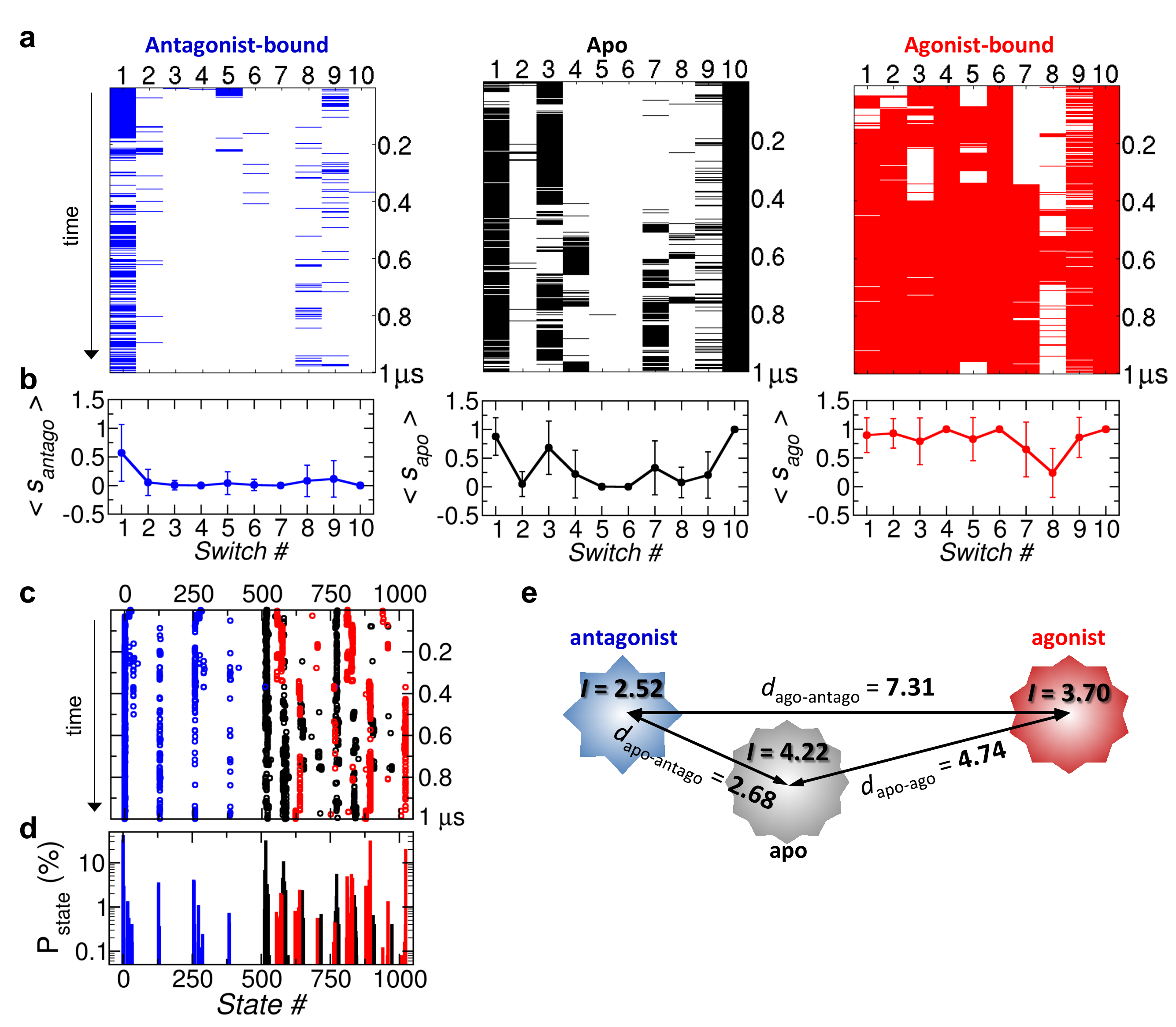}
  \caption{{\bf Dynamics of A$_{2A}$AR in 10-binary switch representation.} 
  (a) Simulation trajectories of antagonist (blue), apo (black), agonist (red) form represented in terms of the ON/OFF state of 10 switches. 
  The lines denote the ON states, and the trajectories evolve from the top to bottom.  
  (b) Mean value of each switch with error bar denoting the standard deviation. 
  (c) Time traces of the microstates represented by the decimal numbers from 0 to 1023 in the apo (black), antagonist-bound (blue), and agonist-bound (red) forms.
  (d) Corresponding population of the microstates. 
  (e) A schematic of similarity between three receptor states in terms of Hamming distance $d_{\alpha\beta}$ with the measure of complexity, $I$ (Eq.~\ref{complexity}), illustrated with polygons.  
 \label{Dynamics_microstates}}
\end{figure}

\begin{figure*}[h]
\centering
  \includegraphics[width=0.9\columnwidth]{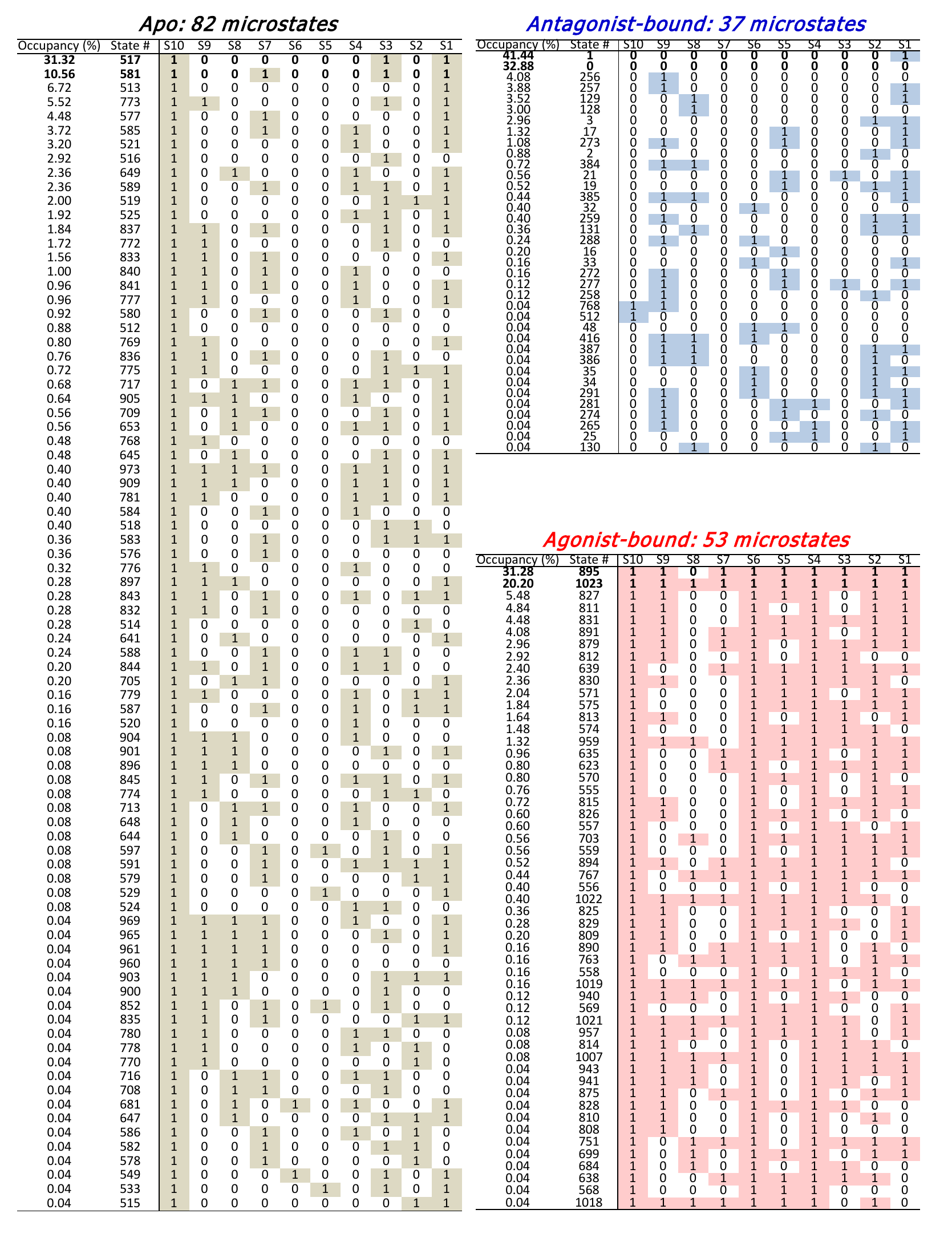}
  \caption{{\bf Microstates observed during the MD simulation and their occupancies.} For each microstate, the switches in the ON state ($s_i=1$) are marked with colored boxes. 
\label{Microstates_statistics}}
\end{figure*}

Representing GPCR conformation in terms of the 10 binary switches amounts to ``choosing" multiple progress coordinates (or multi-dimensional order parameters) to probe the allosteric dynamics of GPCR from the inactive to active state.  
The assumption that 10 binary switch can faithfully represent the dynamics of GPCRs leads to in total, $2^{10}$ possible microstates; each microstate is expressed using binary number from 
0000000000$_{(2)}$ to 1111111111$_{(2)}$ with each digit denoting the switch number from 1 to 10.
These binary numbers can also be expressed with a decimal number from 0 and 1023 (Fig. S6). 
The time traces projected on the 10 binary switches and $2^{10}$ microstates are shown in Fig.~\ref{Dynamics_microstates}a and Fig.~\ref{Dynamics_microstates}c, respectively.  
The average value of each switch, calculated in different receptor state $\xi$ as $0\leq \langle s_{\xi,i}\rangle\leq 1$ (Fig.~\ref{Dynamics_microstates}b), where $\langle s_{\xi,i}\rangle$ with $i$ denoting the switch index is the value of switch averaged over the simulation time, indicates that on average switches are ON in the agonist-bound form, OFF in the antagonist-bound form, and they lie in between in the apo form. 
The difference among the three receptor forms becomes more evident in terms of the population of microstates (Fig.~\ref{Dynamics_microstates}d).
The statistics of microstates shows that the receptor occupies different population of microstates depending on the type of ligand (Fig.~\ref{Dynamics_microstates}).
Among the entire microstates as summarized in Fig.~\ref{Microstates_statistics}, 
(i) $\approx$ 80 \% of antagonist-bound form are populated in the 0000000000 or 0000000001 state. 
(ii) $\approx$ 20.2 \% of the switches in the agonist-bound form are in 1023th state (1111111111$_{(2)}$), and $\approx 31.28 $ \% are in 895th state (1101111111$_{(2)}$). 
(iii) Lastly, in the apo form of GPCR, on an average, 
$\mathcal{S}1$, $\mathcal{S}3$, $\mathcal{S}10$ are ON state, while $\mathcal{S}2$, $\mathcal{S}4$, $\mathcal{S}6$, $\mathcal{S}8$, $\mathcal{S}9$ are OFF state. 
Microstates that constitute the major population of the apo form are 1000000101$_{(2)}$ (31.32 \%) and 1001000101$_{(2)}$ (10.56 \%).

To quantify the statistical similarity explored by two different receptor states, say $\alpha$ and $\beta$ ($\alpha\neq\beta$) in the 10-switch representation, we employ Hamming distance:  
\begin{equation}
d_{\alpha\beta}=\sum_{i=1}^{10} |\langle s_{\alpha,i}\rangle-\langle s_{\beta,i}\rangle|
\label{Hamming}
\end{equation}
where $|x|$ is the absolute value (or modulus) of $x$, and $\langle s_{\xi,i}\rangle$ is the average value of $i$-th switch ($i=1,2\ldots,10$) in the receptor form $\xi$, which is calculated in Fig.~\ref{Dynamics_microstates}b.  
Since $0\leq\langle s_{\xi,i}\rangle\leq 1$, it is expected that $0\leq d_{\alpha\beta}\leq 10$. 
The more similar, the smaller the value of $d_{\alpha\beta}$ should be.       
We obtain $d_{\text{apo-ago}}=4.74$, $d_{\text{apo-antago}}=2.68$, $d_{\text{ago-antago}}=7.31$.

Next, the ``complexity" of each macrostate, defined with the ensemble of microstates $\{i=1\ldots N_s\}$ where $N_s=2^{10}$, is quantified using the Shannon entropy: 
\begin{equation}
I=-\sum_{i=1}^{N_s} p_i\log_2{p_i}
\label{complexity}
\end{equation} 
where $p_i$ is the probability of occupying the $i$-th microstate as in Fig.~\ref{Microstates_statistics}. 
When the receptor explores only a single state ($p_k=1$ and $p_{i\neq k}=0$), the value of $I$ should be $I=I^{\text{min}}=0$; and if all the $2^{10}$ states are uniformly explored ($p_k=2^{-10}$ for all $k$) then $I=I^{\text{max}}=10$. Thus, the larger the value of $I$, the more diverse microstates are explored, which is also gleaned from Fig.~\ref{Microstates_statistics}. 
We obtain $I_{\text{antago}}\approx 2.52$, $I_{\text{apo}}\approx 4.22$, $I_{\text{ago}}\approx 3.70$, for antagonist-bound, apo, and agonist-bound state, resepctively; and hence the apo state explores the most diverse configurational space. 

Fig.~\ref{Dynamics_microstates}e shows a schematic of relationship between the three receptor states combining the analyses using Eq.~\ref{Hamming} and Eq.~\ref{complexity}. 
The apo form is more similar to the antagonist-bound form than to the agonist-bound form, which is consistent with the general notion that agonist contributes to the active GPCRs while both apo and antagonist contribute to the inactive state.   
\\

\begin{figure}[h]
\centering
  \includegraphics[width=0.8\columnwidth]{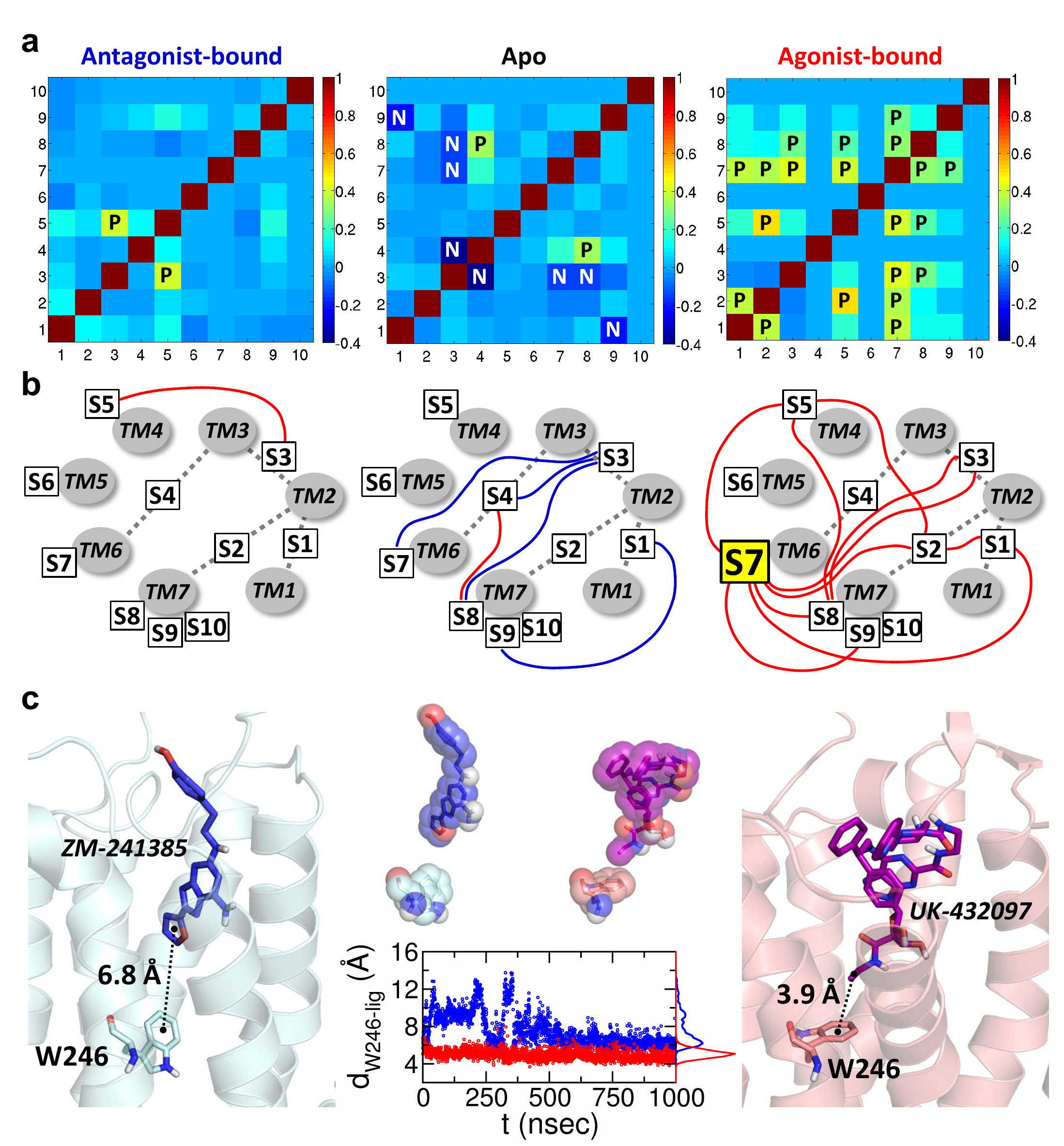}
  \caption{{\bf Cross-correlations among binary switches.}
  (a)  Cross-correlation matrices between the changes in 10 ON/OFF switches for three distinct receptor states calculated by using Fig.~\ref{eqn:covariance}. 
  The symbols ``P" in the matrix elements are for the postive correlatin ($C_{ij}>0.25$); 
  ``N" is for the negative correlation ($C_{ij}<-0.25$). 
   (b) Diagram of the cross-correlation between the switches. TM1 to TM7 helices are displayed in gray circles, and the ten switches are specified with the boxes. 
   The positive and negative correlations are depicted using red and blue lines, respectively.
   (c) Coordination of the antagonist and agonist to $\mathcal{S}7$ (W246).   
   W246 and the bound ligands are depicted in the left and right figures. 
   The graph in the middle shows the distances between the center of mass of the W246 (indole 6-ring) and the center of mass of the furan ring (ZM-241385, blue) and ethyl group (UK-432097, red).
\label{cross_corr}}
\end{figure}

\noindent{\bf Cross-correlations of the dynamics between binary switches.}
To identify the correlation between the ON/OFF dynamics of binary switches, we calculated their cross-correlation ($C_{ij}$) by using the conformational ensemble from the simulations. 
\begin{equation}
C_{ij}=\frac{\langle \delta s_i\delta s_j\rangle}{\sqrt{\langle(\delta s_i)^2\rangle}\sqrt{\langle(\delta s_j)^2\rangle}}
\label{eqn:covariance}
\end{equation} 
where $\delta s_i=s_i-\langle s_i\rangle$ is the variation of the switch value from its mean. 
$C_{ij}$ assesses the extent of coherence in the ``change" in switch dynamics between the $i$-th and $j$-th switches.  
Marked differences of the correlation pattern are observed in the three distinct receptor states 
(Fig.~\ref{cross_corr}a):  
(i) The apo state (the middle panel in Fig.~\ref{cross_corr}) has only one positive correlation $(C_{\mathcal{S}4\mathcal{S}8}>0.25)$, and many other negative correlations $(C_{\mathcal{S}3\mathcal{S}4}, C_{\mathcal{S}3\mathcal{S}7}, C_{\mathcal{S}3\mathcal{S}8}, C_{\mathcal{S}1\mathcal{S}9}<-0.25)$; (ii)  
By contrast, in the antagonist-bound state (the left panel in Fig.~\ref{cross_corr}), a positive correlation $(>0.25)$ is detected only between the $\mathcal{S}3$ and $\mathcal{S}5$, and the negative correlations present in the apo state are suppressed; 
(iii) The agonist-bound state has a greater number of the positive correlations between the switches. 
The diagrams in Fig.~\ref{cross_corr}b illustrate how the allosteric couplings are established among the switches, especially highlighting many positive couplings among switches in the agonist-bound state. 
Most notably, $\mathcal{S}7$ (W246), a central rotameric switch located at the deep bottom of ligand binding cleft, displays direct couplings with 6 other switches $\mathcal{S}1$, $\mathcal{S}2$, $\mathcal{S}3$, $\mathcal{S}5$, $\mathcal{S}8$, and $\mathcal{S}9$, and additional positive correlations are observed in $C_{\mathcal{S}1\mathcal{S}2}$, $C_{\mathcal{S}2\mathcal{S}5}$, $C_{\mathcal{S}3\mathcal{S}8}$, $C_{\mathcal{S}5\mathcal{S}8}$. 
This suggests that intramolecular signaling over the entire structure can be initiated by stimulating the $\mathcal{S}7$.  

As can be confirmed from simulations, the indole 6-ring of W246 is within the range of hydrophobic interaction (4--5 \AA) \cite{Bissantz10JMC} with the ethyl group of the agonist (UK-432097), while such direct interaction with W246 is missing in the antagonist (ZM-241385) (see Fig.~\ref{cross_corr}c). 
Furthermore, there is a marked difference between the antagonist and agonist configurations in the orthosteric binding site; the antagonist is not stably poised as the agonist in the binding cleft (Fig.~\ref{cross_corr}c, middle panel and Fig. S7). 
The dispersion of the center of position for each ligand is $\sigma_{antago}=2.204$ \AA\ and $\sigma_{ago}=0.854$ \AA\ (Fig. S7).     
Although the residues at the binding site adopt diverse configurations, some key residues retain persistent interaction with the receptor (Fig. S8a). Both the antagonist and agonist generally maintain the H-bondings with N253 and E169 residues (Fig. S8b). Also, the adenine rings of the ligands interact with the phenyl ring of F168 via $\pi$-$\pi$ interaction. Fig. S8c shows that the adenyl ring of the agonist interact directly with F168 while that of antagonist does not. 
Occasionally, the antagonist spins or flips at the binding site, indicating the lack of interaction with F168 residue. The polar residues located at the bottom of the binding pocket, i.e., T88, S277, and H278, only interact with the agonist. Especially, T88, one of the hot spot residues in our earlier work \cite{Lee14Proteins}, maintains the stable H-bonding with the agonist.

Along with stable configuration of the agonist ligand (UK-432097) in the binding cleft (See Fig. S7), the widespread cross-correlation among the switch dynamics of $\mathcal{S}1$, $\mathcal{S}2$, $\mathcal{S}3$, $\mathcal{S}5$, $\mathcal{S}7$, $\mathcal{S}8$, and $\mathcal{S}9$ in the agonist form suggests that the interaction between agonist and W246 actuates allosteric signaling and contributes to a stable activation of the receptor function. 
\\

\begin{figure}[h]
\centering
  \includegraphics[width=1.0\columnwidth]{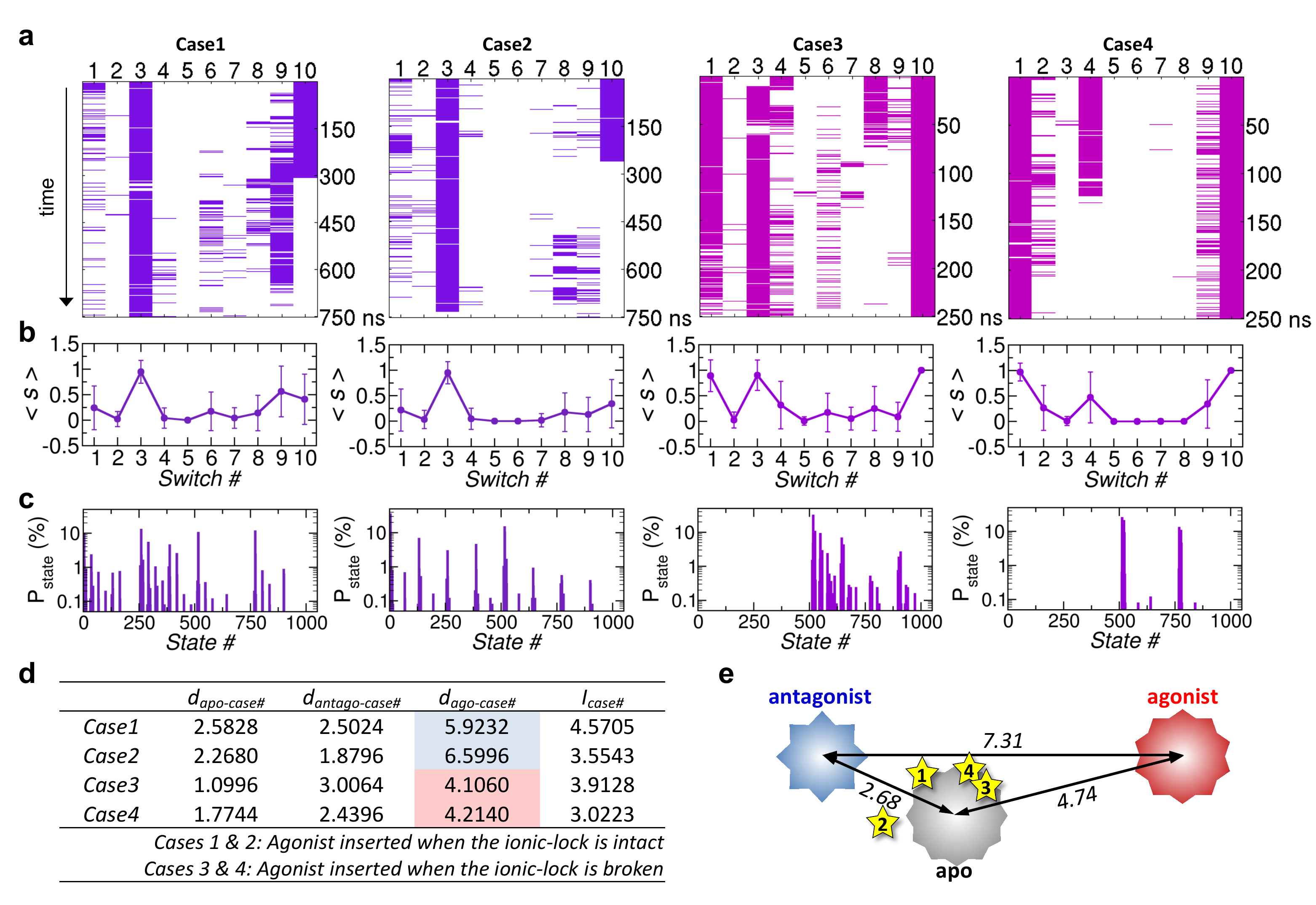}
  \caption{{\bf Agonist inserted to apo form.} 
  (a) Time traces from the case 1 to case 4. In the cases 1 and 2, the agonist was inserted into the apo form when the ionic-lock was intact; whereas in the cases 3 and 4, the agonist was inserted when the ionic-lock was disrupted.   
  (b) Average values of switch from $\mathcal{S}1$ to $\mathcal{S}10$ for the case 1 through 4.  
  (c) Population of microstates sampled after the insertion of agonist. 
  (d) Hamming distance and complexity calculated for cases 1--4.
  (e) The stars are the locations of the cases from 1 to 4, calculated in terms of Hamming distance relative to the apo, antagonist, and agonist forms.  
\label{Ago_to_apo}}
\end{figure}

\noindent{\bf Effect of inserting agonist to the apo state.}
Anticipating a detectable conformational change from inactive to active state, we generated 4 additional time traces by inserting agonist to the orthosteric binding site in the simulation trajectories of the apo state (Fig.~\ref{Ago_to_apo}a, Fig. S9. See Methods for the details of the simulation). 
The first two traces (cases 1 \& 2) were generated by inserting agonist at 125 ns and 150 ns when the ionic-lock was still intact ($s_4=0$) and were simulated for $\approx 750$ ns. 
The second two traces (cases 3 \& 4) were generated at 595 ns and 625 ns when the ionic-lock was disrupted ($s_4=1$), and were simulated for $\approx 250$ ns.
The consequences of the insertion of agonist, summarized in Fig.~\ref{Ago_to_apo}, is still minor, which is evident when the average value of each switch $\langle s_{i}\rangle$ ($i=1,\ldots,10$) is compared (Fig.~\ref{Dynamics_microstates}b middle panel versus Fig.~\ref{Ago_to_apo}b). 
This is not so surprising given that time scale of our simulation ($< 1$ $\mu$sec) is still too short to see a complete transition from inactive to active state in GPCRs, which is typically longer than milisecond time scale \cite{Nygaard2013Cell}.  
Although the overall trend of $\langle s_i\rangle$ looks similar, each trace from the insertion of agonist explores distinct microstate population (Fig.~\ref{Ago_to_apo}c).

Compared with the cases 1 \& 2, 
the 10 switch states were closer to those of agonist-state in the cases 3 \& 4 when the agonist was inserted to the receptor with disrupted ionic-lock ($s_4=1$); in particular, 
the value of $s_4$ for the cases 3 \& 4 is greater and more variable (Fig.~\ref{Ago_to_apo}a,b) although the case 4 shows the stable rebinding of ionic-lock after 120 ns ($s_4=1\rightarrow 0$, the rightmost panel in Fig.~\ref{Ago_to_apo}a). 
The complexity (Eq.\ref{complexity}) of microstate ensemble explored in each simulation, are given in the table of Fig.~\ref{Ago_to_apo}d and the Hamming distances of the cases 1--4 from the three macrostates are calculated in Fig.~\ref{Ago_to_apo}e.  
The cases 3 and 4 are closer to the agoinist-bound state with $d_{\text{ago-case3}}=4.11$ and $d_{\text{ago-case4}}=4.21$ (Fig.~\ref{Ago_to_apo}d, e) than the cases 1 and 2 ($d_{\text{ago-case1}}=5.92$, $d_{\text{ago-case1}}=6.60$).    
It is noteworthy that the binding of agonist to a receptor with stable ionic-lock (cases 1 \& 2) has driven the ensemble of microstates away from agonist-bound form and bring the ensemble close to the antagonist-bound state. 
Disruption of the ionic-lock in DRY motif is required for the activation of A$_{2A}$AR.

\section*{Discussion}
Given that typical time scale associated with the transition from inactive to active GPCR conformers is $\gtrsim\mathcal{O}(1)$ msec \cite{vilardaga2003NatBioTech}, it is practically impossible to capture the complete evidence of transition from inactive to active state using a single time trace lasting only 1 $\mu$sec \cite{Grossfield07Proteins}.
Although the recent improvement in computational power and various computational strategies have ameliorated this time scale gap and have played significant roles in elucidating new facets of GPCR dynamics \cite{Niesen11JACS,kohlhoff2014NatureChem,Dror2011PNAS}, the gap of time scale between all-atom molecular dynamics simulation and experiments is still a serious problem in linking the computation of biomolecular dynamics with experimental observation \cite{hyeon2011NatureComm}.
Rather short in time scale ($\sim$ 1 $\mu$sec), 
the trajectories from our simulations, which essentially probe the dynamics of receptor in terms of the noisy local variables, enable us to map the intramolecular signaling of three different ligand states. 
When it comes to the sampling efficiency of each macrostate, 1 $\mu$sec time scale is not too short to observe the individual dynamics of side chain flips that generally occur in picoseconds to nanoseconds timescale \cite{Dror2012ARB}, and we tried to decipher collective signals out of those noisy individual signals. 
The dynamic characteristics of the three macrostates are well discerned in consistent with the general notion of GPCRs, and our analysis based on the dynamics of 10 binary switches quantifies the differences among the thre statese. 

Our simulation results provide a comprehensive view of ligand-dependent conformational dynamics and show how the remote allosteric hotspots of A$_{2A}$AR are coupled each other, which lend support on the pre-existing experimental data.
For example, W246 has long been proposed to function as a central rotamer switch in the activation of GPCRs \cite{nygaard2009TPS}. 
While the rotameric change of W246 was confirmed in spectroscopy \cite{Crocker2006JMB} and some MD simulation studies \cite{nygaard2009TPS,Li2013JACS}, 
X-ray crystal structures of A$_{2A}$AR in both antagonist-bound and agonist-bound states show little difference of the position and dihedral angles of this residue. 
Our simulation captures the transition of the indole ring of W246 in the presence of agonist (Supporting movie M1). 
In addition, our simulations clearly demonstrate characteristics of local dynamics of microswitches such as ligand-dependent state of ionic-lock, and the ligand-dependent variation of the rotameric angles in Y197 and Y288 \cite{jaakola2008Science, xu2011Science, lebon2011Nature, Dore2011Structure}. 

%For example, using cloud-based computing, Kohlhoff \emph{et al.} have stitched thousands of short time traces into $\sim 150$-$\mu$sec time traces by assuming Markovian processes between discrete conformational states, and studied the pathways of $\beta_2$ adrenergic receptor \cite{kohlhoff2014NatureChem}. 
%Dror \emph{et al.} have generated ensemble of unbiased MD simulations, each of which was extended to $\approx$ 10 $\mu$sec, and monitored the ligand binding induced activation or inactivation of $\beta$2 adrenergic receptor \cite{Dror2011PNAS} as well as the process of ligand binding to the orthosteric site \cite{dror2011PNAS1}. 
%Similar to our study that has elucidated the correlations between binary switches, they drew a conclusion that a loosely coupled allosteric networks among three different regions in structure is a basis of allosteric response and signaling for the receptor activation \cite{Dror2011PNAS}. 

Until recently, there are some other studies that have compared the conformational differences of A$_{2A}$AR depending on the receptor states \cite{Li2013JACS,Pang2013Proteins,Lee2013CPL}. 
Pang \emph{et al.} studied dynamic behaviors of A$_{2A}$AR induced by the binding of two distinct antagonists \cite{Pang2013Proteins}. 
Lee \emph{et al.} simulated the ligand-dependent cholesterol interactions in A$_{2A}$AR \cite{Lee2013CPL}. 
Exploring the distinct structural states that resemble the active and inactive states, Li \emph{et al.} observed that the key structural elements change in a highly concerted fashion during the conformational transition \cite{Li2013JACS}. 
Similar to our study, Li \emph{et al.} highlighted the importance of the rotameric transition of W246 during the activation process. 
In contrast to these studies, which focused more on the ligand-receptor interactions, we tried to identify the communication among the microswitches. 
To best of our knowledge, our simulation is the first systematic study probing the dynamics of all microswtich residues of A$_{2A}$AR, and made explicit that there are marked differences among the ligand-dependent conformational ensembles; and thus each ligand-dependent macrostate is made of mutually exclusive population of microstates, which supports the view that different type of ligands effectively remodel the protein energy landscapes \cite{Motlagh2014Nature,Smock2009Science}. 

In describing the conformational transition from inactive to active states in GPCRs, the recent studies by Yuan \emph{et al.} \cite{Yuan2014NatCommun} and Bhattacharya \emph{et al.} \cite{Bhattacharya2014BiophysJ} showed consistent and complementary results to our work, although the used methodologies were different. 
Yuan \emph{et al.} discovered that a hydrophobic layer located inside of the helix structure forms a gate that opens to form a continuous water channel only upon the agonist binding \cite{Yuan2014NatCommun}. The configuration of the nearby NPxxY motif affects this water channel, and the G$\alpha$ peptide stabilizes the active state of Y7.53 which is one of the 10 switches in our study. Also, they calculated the average volume of the intracellular G-protein binding site and found that the volume is significantly increased upon the activation process, so that the intracellular site can accommodate G-protein binding. The volume change might well be a consequence of the swinging motion of TM5-ICL3-TM6, and consistent with our result showing that the distance between the intracellular loop 2 (ICL2) and ICL3 varies depending on the ligand binding states. 
Our study revealed the correlation between the ICL3 motion and the ionic-lock formation (Fig.~\ref{DRY_motif}b).
Bhattacharya \emph{et al.} calculated the mutual information (MI) in internal coordinates from MD simulated trajectories, and they identified allosteric communication pipelines which are cenceptually similar to the long-range cross-correlation pathways discussed in our previous work \cite{Lee14Proteins}. They found that, in the inactive state, the allosteric pipelines mainly cross the TM6 helix, and as the state moves from intermediate to agonist-bound active state, these pipelines pass though the TM7 helix, suggesting that TM7 is increasingly correlated in the active state.
As suggested in our study, diverse conformational space of GPCRs is dependent on the ligand binding states and is regulated by the allosteric pathways comprising of some hub residues. 
The works by these two groups and ours both surmise TM7 as the key helix in GPCR activation. 
While the most dynamic part is TM5-ICL3-TM6 region which accompanies the swinging motion, the hub residues, scattered throughout the helices, regulate the activation process in a concerted way.

The antagonist-bound form explores the rotamer angle space entirely different from those in the agonist-bound and apo forms.
The microstates visited by the apo form show an overlap with those by agonist-bound form although the extent of such overlap in the entire microstate population is vanishingly small and the duration of such overlap is only transient \cite{Yao2009PNAS}.
Occasional 10$^o$-tilt of TM5-ICL3-TM6 relative to TM3, shown in the simulation of the apo state (Fig.~\ref{DRY_motif}b), may be related to the basal activities of the GPCRs.
According to the pre-coupled theory, several class A GPCRs are suspected to be coupled to their corresponding G-proteins even in the inactive or basal states \cite{Gales2006NSMB,Philip2007JBC,Qin2011NCB,Jakubik2011PlosOne}. 
Inactive-state preassembly can facilitate the rapid and specific G protein activation\cite{Qin2011NCB}.
Our simulation results suggest that the apo state of A$_{2A}$AR can also form an inactive-state preassembly by visiting the microstates that overlap with those of the antagonist-bound state. 
We expect that a simulation of the apo form conducted in the presence of a G-protein will amplify this overlap and change the dynamics and correlation between the hotspots as well.

Among several findings and predictions made in the present study, of particular note is the role of W246 ($\mathcal{S}7$) in the GPCR activation. Although the importance of W246 residue were largely documented based on experimental or theoretical studies\cite{nygaard2009TPS,Crocker2006JMB,Li2013JACS}, our cross-correlation analysis unequivocally indicates that W246 can work as a hub in the communications among the important microswitch residues (Fig.~\ref{cross_corr}). 
Positioned deep inside the binding cleft, a signal of rotameric change of W246, triggered by a direct hydrophobic interaction with an agonist, can be transmitted to the 6 other switches ($\mathcal{S}$1, $\mathcal{S}$2, $\mathcal{S}$3, $\mathcal{S}$5, $\mathcal{S}$8, $\mathcal{S}$9) that W246 is in direct correlation with. 
Our study confirms that W246 plays pivotal roles in GPCR activation as both an agonist sensor and actuator of allosteric signaling.

The class A GPCRs are conventionally reported to function as monomers\cite{Maurice2011TPS,Ernst2007PNAS,Filizola2010LifeSci,Whorton2007PNAS}; however,
growing body of experimental evidence indicates that some GPCRs, including A$_{2A}$AR, can form homodimers, heterodimers, or even higher level of oligomers\cite{Terrillon2004EMBO,Vidi2008FEBS,Fanelli2011BiochimBiophysActa}.
In recent years, single-molecule imaging techniques revealed that GPCRs undergo dynamic equilibrium between monomers and dimers \cite{Kasai2014CurrOpinCellBiol}, and studies of GPCR monomers and dimers are both meaningful and necessary. 
The dynamic features and correlations revealed in this study for monomer of A$_{2A}$AR could  change when the receptor forms dimers or higher level of oligomers. 
Thus, it would be interesting to investigate how dimerization of GPCR alters the correlations between hotspots.

Faithful description of biomolecular dynamics, as a complex system, is a major challenge in both computational and experimental molecular biology. In this study, we try to simplify the description of each ligand-bound macrostate by “selecting” the 10 binary switches as the reaction coordinates. 
The conformational features of A$_{2A}$AR captured in our 10 binary switch representation confirmed existing knowledge on the receptor and made specific predictions amenable to a further experimental study.

%\section{  Concluding Remarks}
%As the agonist binding site and the G-protein binding site are separated by TM region, information transfer between these two remote sites is critical for the GPCR activation and regulation. 
%Using A$_{2A}$AR as a model protein, 
%we simulated the dynamics of the GPCR and analyzed how  the conformational space in terms of 10 binary switches is explored depending on the receptor state.  Unbiased MD simulations were carried out for three independent systems, i.e., apo, agonist-bound, and antagonist-bound forms, in an explicit membrane and water environment. 
%We then extensively collected the conformational states of the micro-switch residues and thoroughly investigated their time traces and the cross-correlations. 
%Our extensive conformational analysis of the micro-switch residues highlighted the difference between ligand-dependent states by demonstrating the cross-correlations between binary switches. 
%The information on dynamic couplings between binary switches elucidated here not only provides a global picture of the regulatory mechanism underlying A$_{2A}$AR activation but could also serves as a useful benchmark for the structure-based drug design of its allosteric modulators  \cite{hardy2004COSB,Jensen2004EJPS}. 

% You may title this section "Methods" or "Models". 
% "Models" is not a valid title for PLoS ONE authors. However, PLoS ONE
% authors may use "Analysis" 
\section*{Methods}

{\bf Preparation of the apo, antagonist-, and agonist-bound structures.}
The X-ray crystal structures of A$_{2A}$AR bound with an agonist or antagonist were retrieved from the Protein Data Bank (PDB). Since some loop regions are not resolved in these crystal structures, homology modeling was performed using the MODELLER program implemented in Discovery Studio v.3.1 to prepare the full-length A$_{2A}$AR models including all the loop regions. We used the mutation-free X-ray crystal structures of PDB IDs 3QAK \cite{xu2011Science}  and 3EML \cite{jaakola2008Science}, which were available at the year 2011 we began this study, as the main templates for the agonist-bound and antagonist-bound models, respectively. 
To model the loop regions that were not determined in 3QAK and 3EML, 
the X-ray crystal structures with PDB IDs of 2YDV \cite{lebon2011Nature}  and 3PWH \cite{Dore2011Structure}  were used by retaining the conserved disulfide bridges connecting the loops, i.e., C71-C159, C74-C146, C77-C166, and C259-C262.
These models were optimized by a simulated annealing step and selected based on the Discrete Optimized Protein Energy (DOPE) score \cite{Shen2006ProtSci}. The final structures were energy-minimized with the Conjugate Gradient method and the Generalized Born with simple SWitching (GBSW) implicit solvent model \cite{im2003JCC}. 
We obtained the apo form by minimizing the receptor strucuture after removing the ligand from the antagonist-bound form. 
\\

\noindent{\bf Simulations.}
By employing X-ray crystal structures of A$_{2A}$AR and homology modeling to resolve the missing residues in the loop as described above, 
we performed MD simulations   using NAMD package v.2.8   with   CHARMM22/CMAP   force field. 
 The topology and parameter files for the ligands were generated by SwissParam web server\cite{Zoete2011JComputChem}.

To construct the explicit membrane system, we first predicted the TM region of the A$_{2A}$AR based on the Orientations of Proteins in Membranes (OPM) database, and subsequently surround the TM region with 173 (apo), 182 (antagonist-bound), and 177 (agonist-bound) 1-Palmitoyl-2-oleoylphosphatidylcholine (POPC) lipid molecules, which form a bilayer embedding the receptor (length: 88 \AA\ in x-axis, 91 \AA\ in y-axis). 
The receptor in the membrane system was then solvated with 13,549 (apo), 13,551 (antagonist-bound), and 13,552 (agonist-bound) water molecules. 
We added 38 K$^+$ and 49 Cl$^-$ ions to make $\sim 150$ mM salt condition. 
As a result, the simulation system contains 68,799 (apo), 70,052 (antagonist-bound), and 69,449 (agonist-bound) atoms in 85 \AA$\times$88 \AA$\times$99 \AA\ periodic box. 
Nonbonded interactions were smoothly switched off between 10 and 12 \AA. To handle electrostatic interactions, the particle mesh Ewald algorithm was employed with a grid spacing smaller than 1 \AA. 

The simulations were conducted via (i) energy-minimization of the initial system using the cojugate gradient method  
in the order of membrane, water molecules, and the entire molecules; 
(ii) gradual heating from 0 K to 300 K using a 0.01 K interval at each step; 
(iii) 50 nsec equilibration with NVT ensemble; 
and (iv) $\sim$ 1 $\mu sec$ production run with NPT ensemble for each system with different ligands.  
No specific constraints, such as distance or angle, were applied during the simulations, except SETTLE algorithm for constraints in water molecules.  
We used the integration time step of 1 fsec, and for the analysis saved the simulated trajectories every 2 psec and sampled every 400 psec.

To simulate the effect of inserting the agonist into the simulated apo-state, we first aligned the agonist ligand structure (co-crystallized  conformation in 3QAK.pdb) to the apo-state structure, and then deleted the water molecules around 5 \AA\ from the region where the ligand is to be inserted. After inserting the agonist, we performed the energy minimization and equilibrated the system for 20 nsec, so that water can solvate the ligand binding site as well as the agonist. 
The effect of agonist binding on the receptor structure was monitored subsequently.
\\

% Do NOT remove this, even if you are not including acknowledgments
\section*{Acknowledgments}
This work was partly supported by the grant from the Basic Science Research Program 
(Grant number 2013R1A6A3A01066055) (to Y.L.) and the National Leading Research Lab (NLRL) program (2011-0028885) (to S.C.) funded by the Ministry of Science, ICT \& Future Planning (MSIP) and the National Research Foundation of Korea (NRF).
C.H. thanks the KITP at the University of California, Santa Barbara, for support during the preparation of the manuscript (NSF PHY11-25915).
We thank KIAS and KISTI Supercomputing Center for providing computing resources.

%\section*{References}
% The bibtex filename
\bibliography{mybib1}

\clearpage 
%Fig.1

%Fig.2

%Fig.3

%Fig.4

%Fig.5

%Fig.6

%Fig.7

%Fig.8
\clearpage
\section*{Supporting Figures}
\makeatletter 
\setcounter{figure}{0}
\renewcommand{\thefigure}{S\@arabic\c@figure}
\makeatother 

%Fig.S1
\begin{figure*}[h]
  \includegraphics[width=0.9\columnwidth]{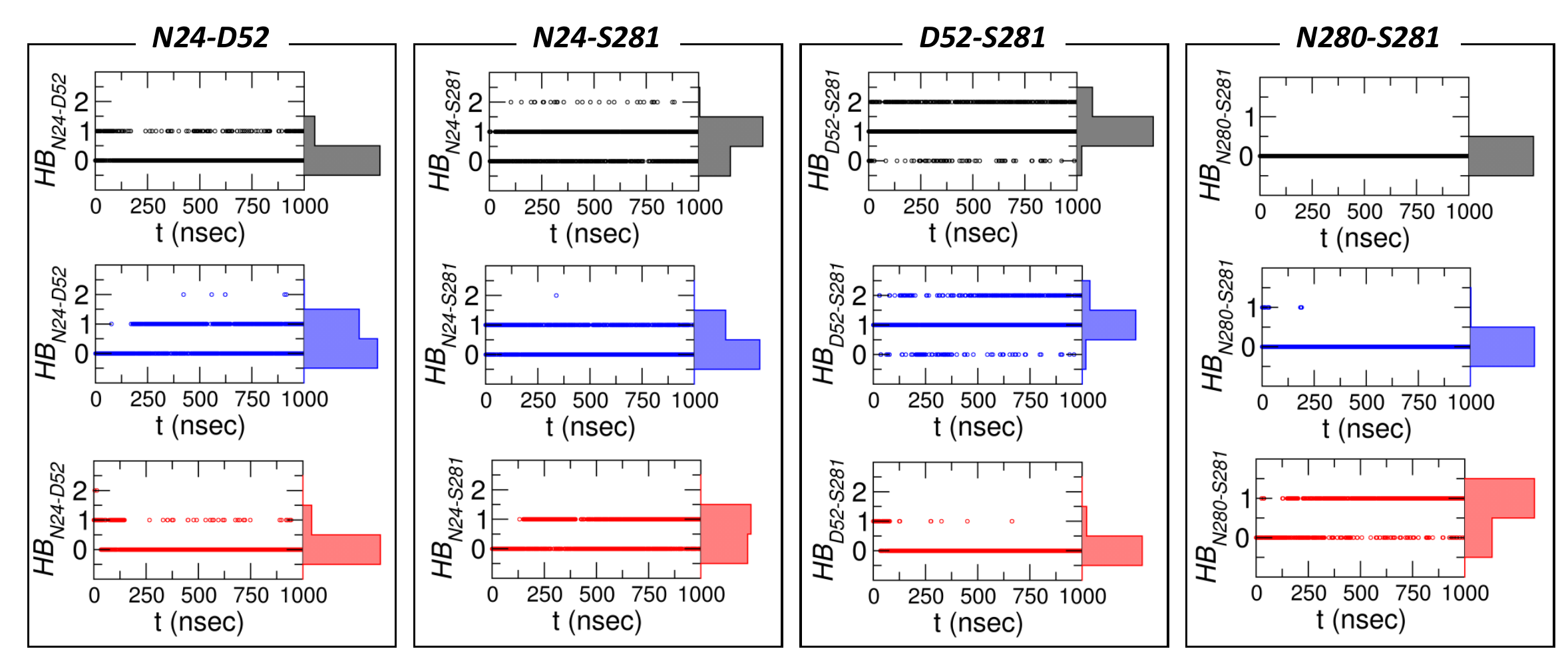}
  \caption{{\bf Hydrogen bonding network of the micro-switch residues in TM1, TM2, and TM7.} The time traces are colored in black, blue, and red for the apo, antagonist-bound, and agonist-bound states, respectively, and their histograms are shown in the right side of the graphs. 
 \label{TM1-2_HB_network}}
\end{figure*}

%Fig.S2
\begin{figure*}[h]
  \includegraphics[width=0.9\columnwidth]{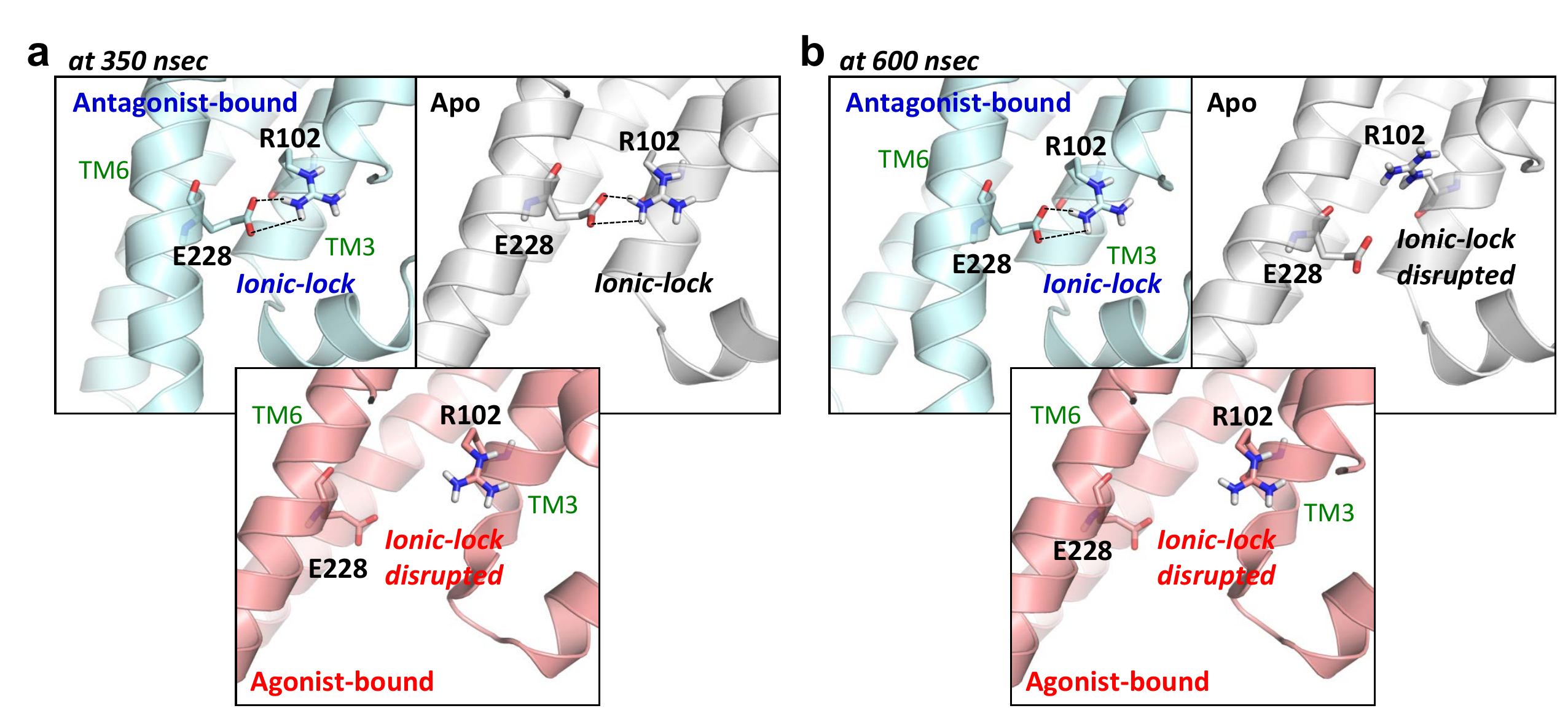}
  \caption{{\bf Comparison of the ionic-lock formation} (a) observed at 350 nsec and (b) at 600 nsec on the left panel of Figure 2b. The ionic-lock was broken at 600 nsec in the apo form.
 \label{ionic_lock}}
\end{figure*}

%Fig.S3
\begin{figure*}[t]
  \includegraphics[width=1.02\columnwidth]{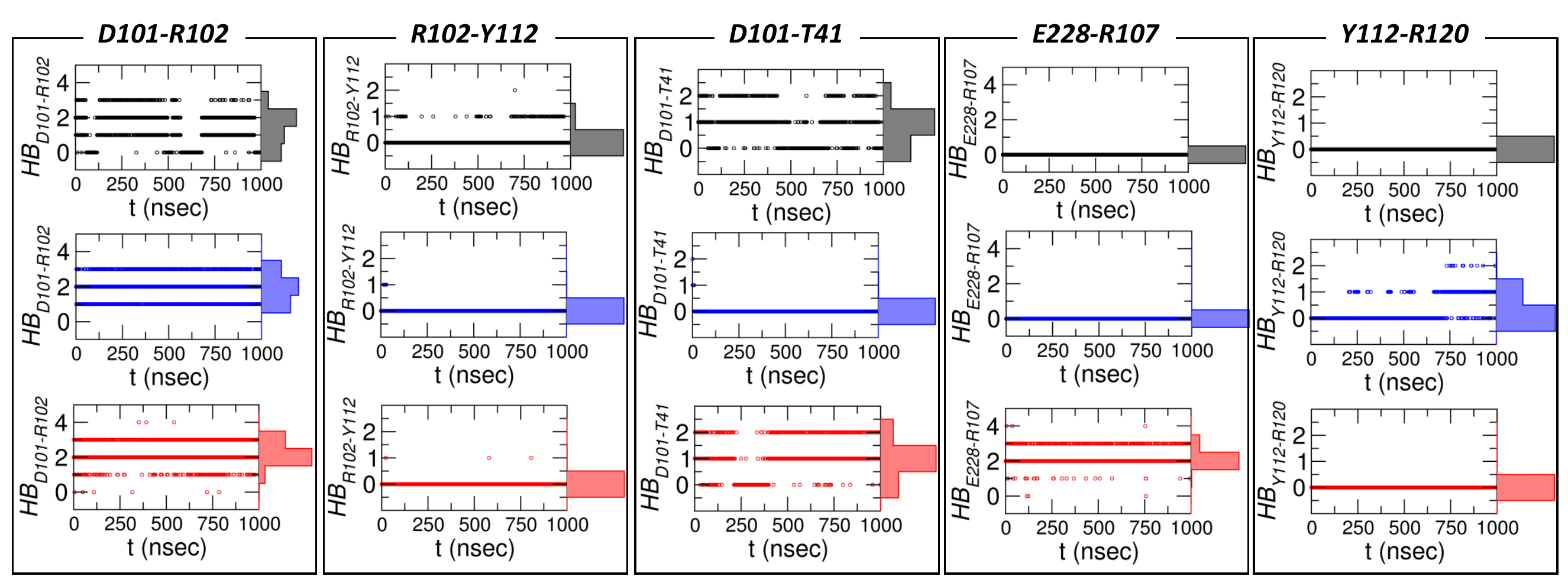}
  \caption{{\bf Hydrogen bonding network of the DRY motif.} The time traces are colored in black, blue, and red for the apo, antagonist-bound, and agonist-bound states, respectively, and their histograms are shown on the right hand side of the traces. 
  \label{DRY_motif_HB_network}}
\end{figure*}

%Fig.S4
\begin{figure*}[t]
\centering
  \includegraphics[width=\columnwidth]{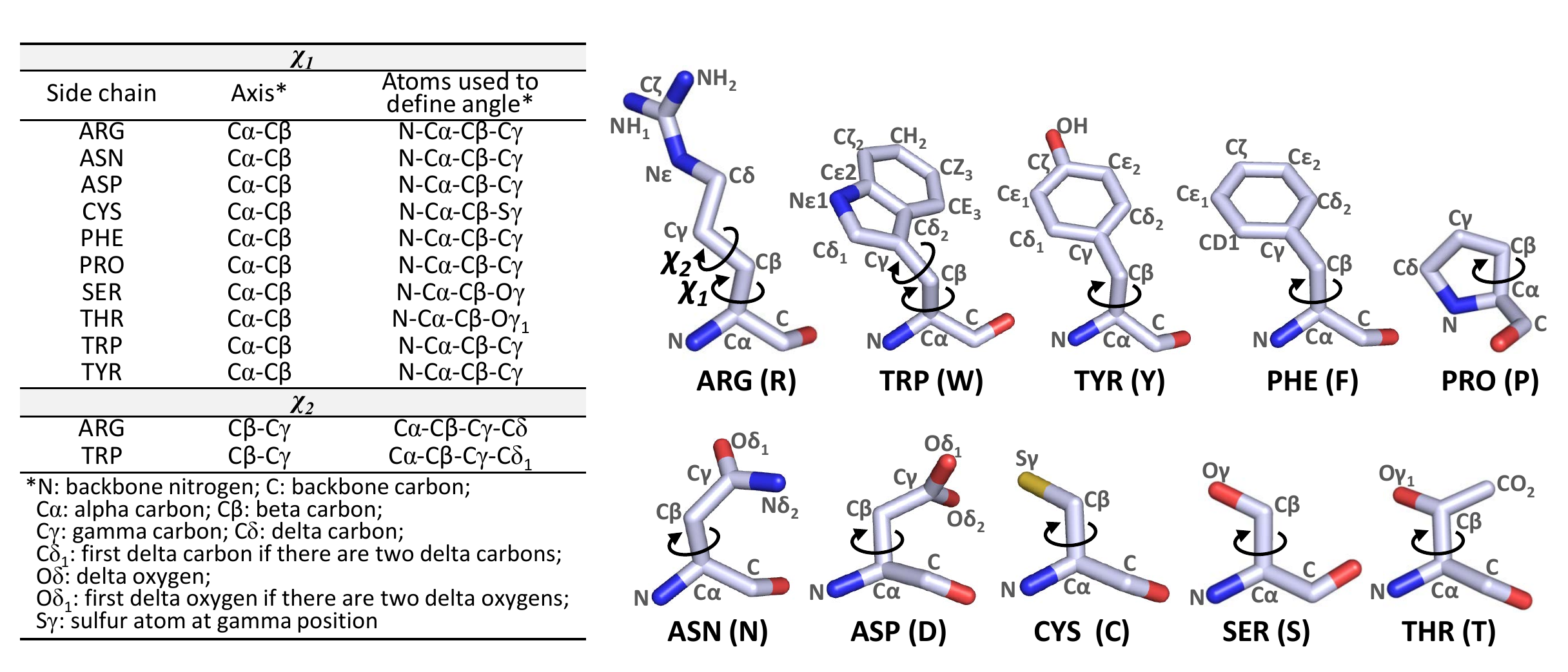}
  \caption{{\bf Dihedral angles that define rotameric changes of the micro-switch residues.} We focused on the $\chi_1$ angles, but also considered the $\chi_2$ angles for Arg and Trp residues.
  \label{dihedral_define}}
\end{figure*}

%Fig.S5
\begin{figure*}[t]
  \includegraphics[width=0.8\columnwidth]{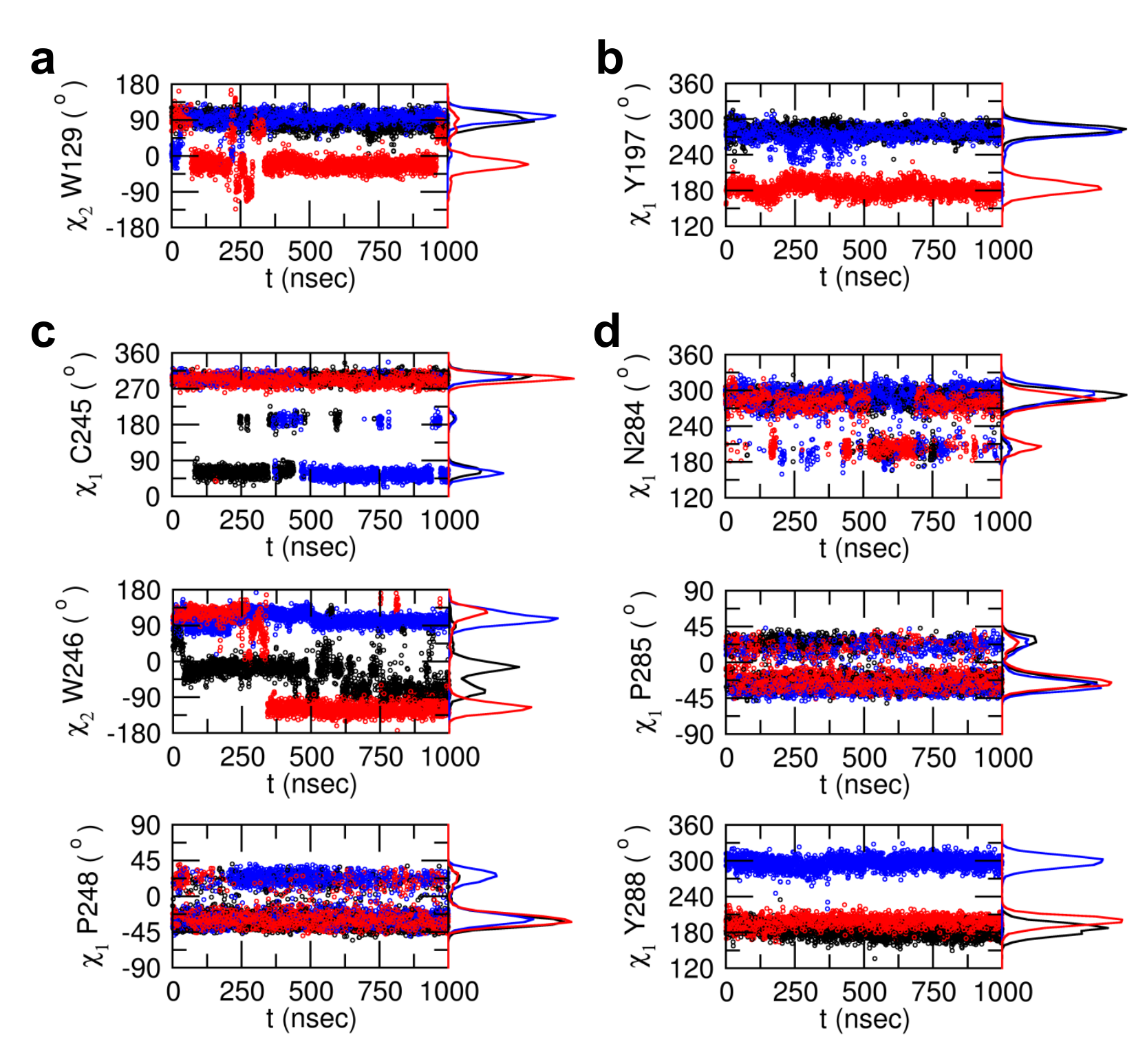}
  \caption{{\bf Time traces of the rotameric changes.} Rotameric states of (a) W129, (b) Y197, (c) CWxP motif, and (d) NPxxY motif are compared. In the graphs, the dihedral angles are colored in black, blue, and red for the apo, antagonist-bound, and agonist-bound states, respectively, and their corresponding histograms are shown on the right side. 
 \label{rotamers_trace}}
\end{figure*}

%Fig.S6
\begin{figure*}[t]
  \includegraphics[width=0.9\columnwidth]{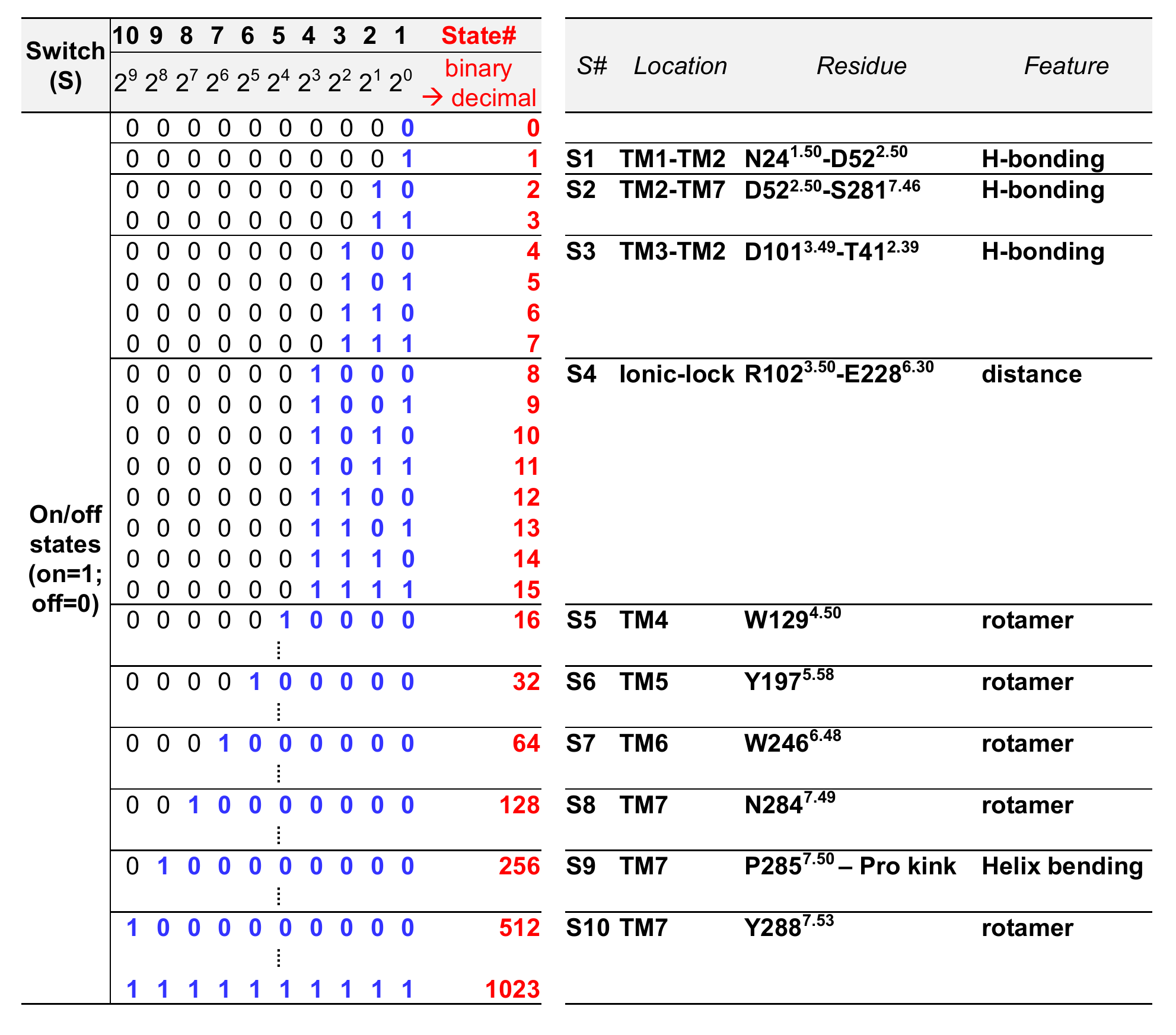}
  \caption{{\bf Indexing rule of the conformational ensembles represented by the ten switches.} If a switch is in the ON state, we defined it as ``1", and in the OFF state, as ``0". The combination of the ON/OFF states made by the ten switches is represented as binary code, and the corresponding state index in decimal number is shown in red.\label{microstates}}
\end{figure*}

%Fig.S7
\begin{figure}[t]
\centering
  \includegraphics[width=0.5\columnwidth]{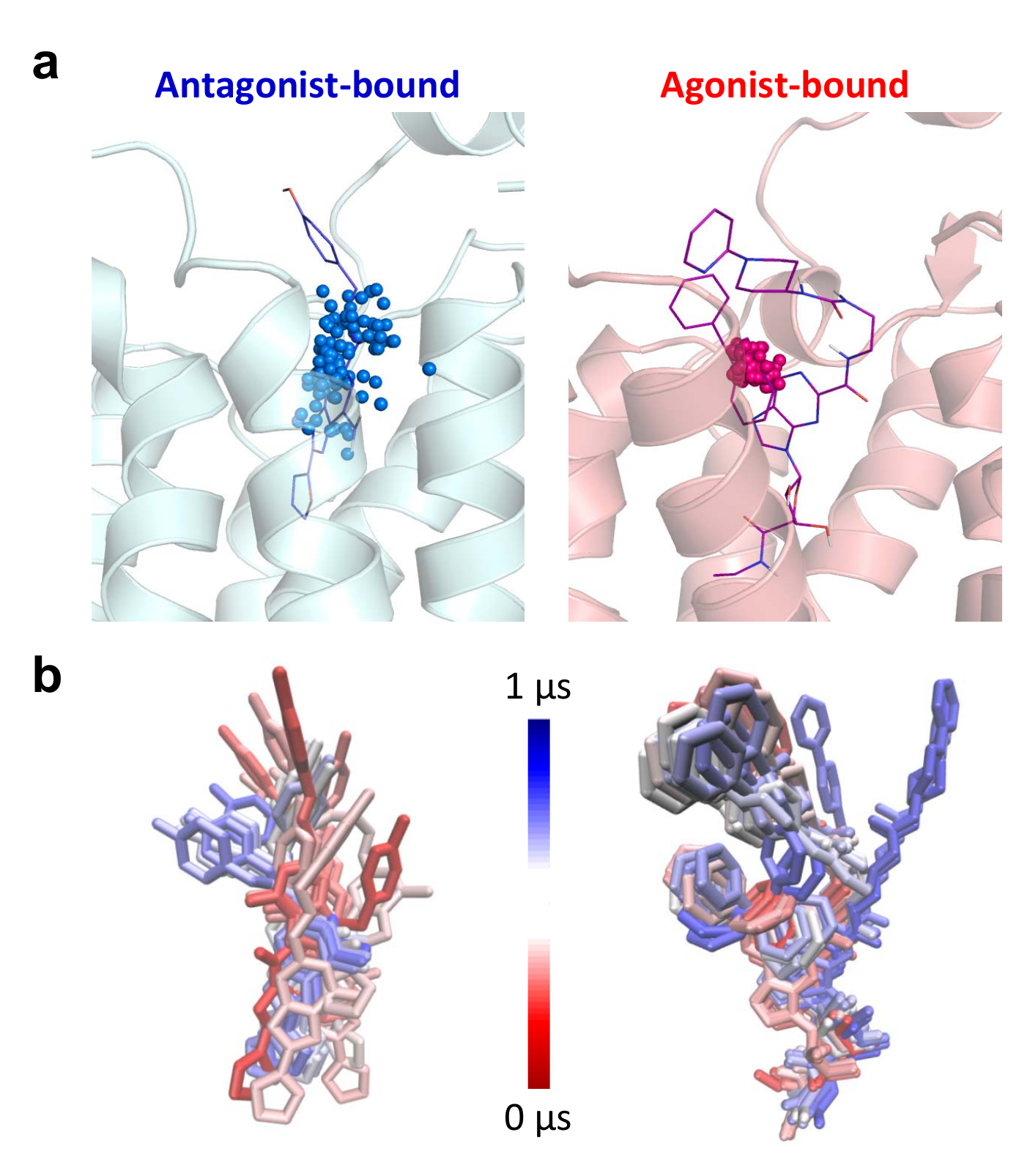}
  \caption{Configurations of antagonist and agonist in the binding cleft during MD simulation. 
  (a) The space sampled by the center of mass of the antagonist and agonist in the binding cleft are depicted with blue and red spheres, respectively. (b) Conformations of the ligands are colored from red to blue for the 1 $\mu$sec simulations \label{ligand_com}}
\end{figure}

%Fig.S8
\begin{figure}[t]
\centering
  \includegraphics[width=0.8\columnwidth]{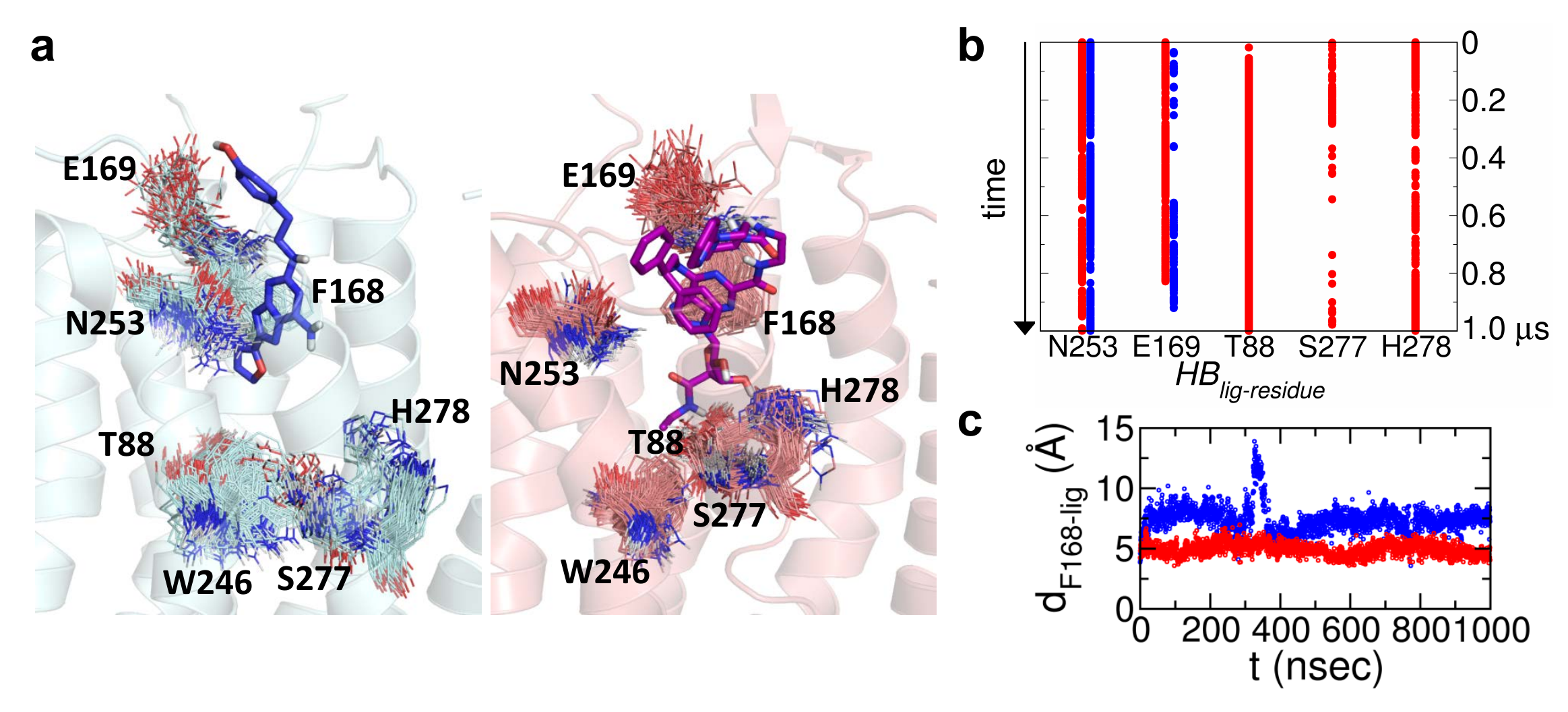}
  \caption{Key interactions of antagonist and agonist in the binding site of hA$_{2A}$AR. (a) Conformations of the interacting residues in the antagonist-bound (left) and agonist-bound (right) forms. (b) Hydrogen bond formation between the residues and antagonist (blue) or agonist (red). (c) Distance between F168 (center of mass of the phenyl ring) and the bound ligands (center of mass of the adenine ring). \label{ligand_interaction}}
\end{figure}

%Fig.S9
\begin{figure*}[t]
\centering
  \includegraphics[width=0.8\columnwidth]{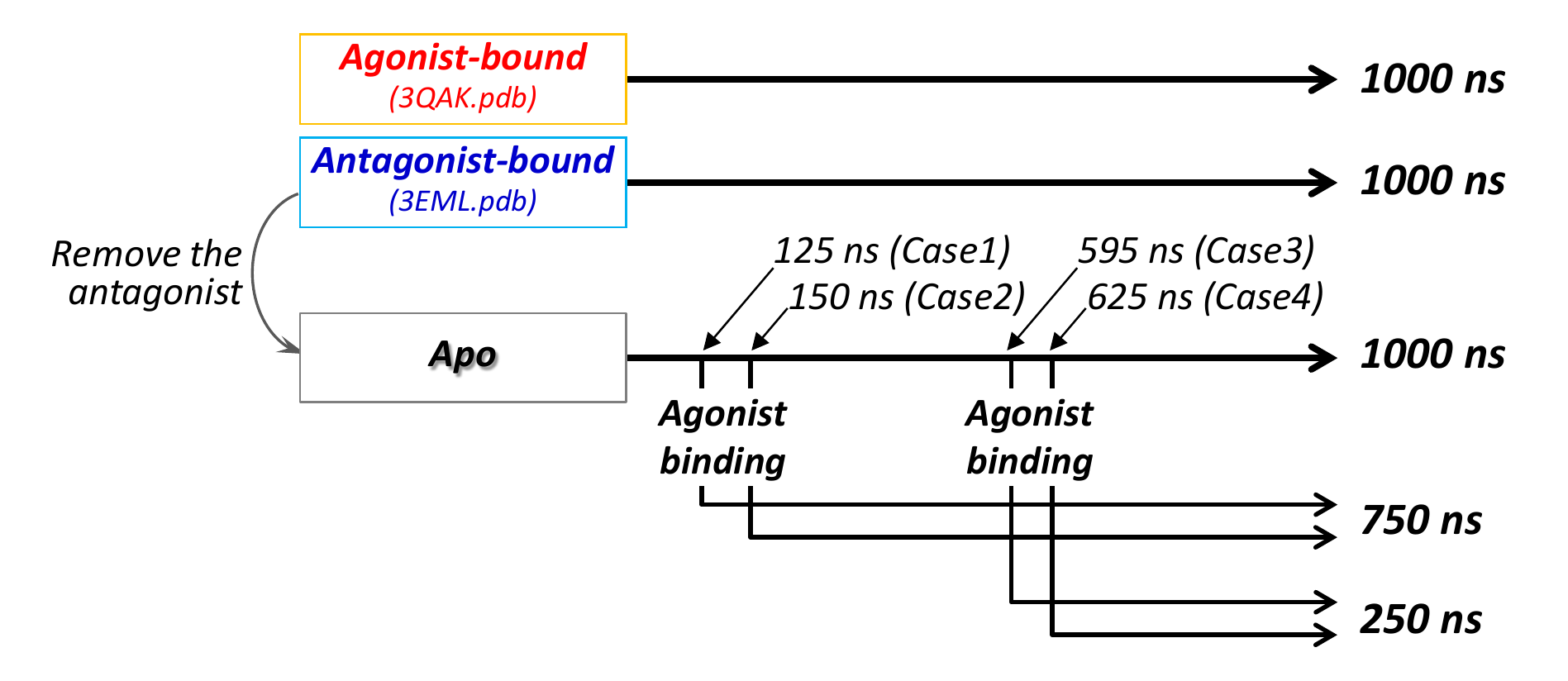}
  \caption{{\bf Insertion of agonist during the simulation of apo state.} 
  We generated four additional time traces by inserting agonist to the simulation trajectory of apo state. 
  The first two traces are generated by inserting agonist at 125 ns and 150 ns of original time trace of apo state in Fig. 5a when the ionic-lock is still intact and simulated for $\approx 750$ ns. The second two traces are generated at 595 ns and 625 ns when the ionic-lock is disrupted, and simulated for $\approx 250$ ns.
  \label{simulation_scheme}}
\end{figure*}

\end{document}